\documentclass[sigconf,table,authorversion=true]{acmart} 

\AtBeginDocument{%
  }

\copyrightyear{2026}
\acmYear{2026}
\setcopyright{cc}
\setcctype{by}
\acmConference[CHI '26]{Proceedings of the 2026 CHI Conference on Human Factors in Computing Systems}{April 13--17, 2026}{Barcelona, Spain}
\acmBooktitle{Proceedings of the 2026 CHI Conference on Human Factors in Computing Systems (CHI '26), April 13--17, 2026, Barcelona, Spain}
\acmPrice{}
\acmDOI{10.1145/3772318.3791912}
\acmISBN{979-8-4007-2278-3/2026/04}

\usepackage{url}
\usepackage{multirow}
\usepackage{subcaption}
\usepackage{fontawesome}
\usepackage{siunitx}
\usepackage{amsmath}
\usepackage{enumitem}
\usepackage{ifthen}
\usepackage{graphicx}
\usepackage{soul}

\newboolean{showrevisions}
\setboolean{showrevisions}{false}
\newboolean{isdraft} 
\setboolean{isdraft}{true} 

\definecolor{contentColor}{RGB}{255, 255, 255}
\definecolor{interactionColor}{RGB}{255, 255, 255}
\definecolor{transitionColor}{RGB}{255, 255, 255}

\newcommand{\preview}{\colorbox{contentColor}{preview}}
\newcommand{\previews}{\colorbox{contentColor}{previews}}
\newcommand{\interaction}{\colorbox{interactionColor}{interaction}}
\newcommand{\interacts}{\colorbox{interactionColor}{interacts}}
\newcommand{\activation}{\colorbox{interactionColor}{activation}}

\newcommand{\selects}{\colorbox{interactionColor}{selects}}

\newcommand{\confirm}{\colorbox{interactionColor}{confirm}}
\newcommand{\transition}{\colorbox{transitionColor}{transition}}
\renewcommand{\preview}{preview}
\renewcommand{\previews}{previews}
\renewcommand{\interaction}{interaction}
\renewcommand{\interacts}{interacts}
\renewcommand{\activation}{activation}

\renewcommand{\selects}{selects}

\renewcommand{\confirm}{confirm}
\renewcommand{\transition}{transition}

\newcommand{\portal}{\textsc{Portal}}
\newcommand{\wim}{\textsc{WiM}}

\newcommand{\gallery}{\textsc{Gallery}}
\newcommand{\headPalette}{\textsc{Head Palette}}
\newcommand{\handPalette}{\textsc{Hand Palette}}
\newcommand{\steeringWheel}{\textsc{World Wheel}}

\newcommand{\techsep}{\hspace{0pt}\discretionary{--}{}{--}\hspace{0pt}}

\newcommand{\baseline}{\textsc{Baseline}}

\newcommand{\portalGallery}{\textsc{Portal}\techsep\textsc{Gallery}}
\newcommand{\portalHandPalette}{\textsc{Portal}\techsep\textsc{HandPalette}}
\newcommand{\portalHeadPalette}{\textsc{Portal}\techsep\textsc{HeadPalette}}
\newcommand{\portalSteeringWheel}{\textsc{Portal}\techsep\textsc{WorldWheel}}

\newcommand{\wimGallery}{\textsc{WiM}\techsep\textsc{Gallery}}
\newcommand{\wimHandPalette}{\textsc{WiM}\techsep\textsc{HandPalette}}
\newcommand{\wimHeadPalette}{\textsc{WiM}\techsep\textsc{HeadPalette}}
\newcommand{\wimSteeringWheel}{\textsc{WiM}\techsep\textsc{WorldWheel}}

\ifthenelse{\boolean{isdraft}}{\newcommand{\todo}[2]{~\\ \textcolor{red}{\textbf{[\textsc{\textcolor{green}{#1}}: #2]}}}}{\newcommand{\todo}[2]{}} 
\ifthenelse{\boolean{isdraft}}{\newcommand{\note}[1]{\textcolor{red}{\textbf{[#1]}}}}{\newcommand{\note}[1]{}} 

\definecolor{newcolor}{rgb}{0.15, 0.7, 0.15}
\definecolor{newcolor}{rgb}{0.1, 0.1, 0.9}
\definecolor{changedcolor}{rgb}{0.1, 0.1, 0.9}
\definecolor{removedcolor}{rgb}{0.9, 0.1, 0.1}

\ifthenelse{\boolean{showrevisions}}{}{}
\ifthenelse{\boolean{showrevisions}}{}{}
\ifthenelse{ \boolean{showrevisions} } 
    {
    \newcommand{\removed}[1]{%
        \textcolor{removedcolor}{\sout{#1}}%
        }
    }
    {
    \newcommand{
        \removed}[1]{}
    }

\ifthenelse{ \boolean{showrevisions} }
    {
        \setlength{\fboxrule}{2pt} 
    }
    {
        \setlength{\fboxrule}{0pt} 
    }

\newcommand{\removefig}[1]{%
    \ifthenelse{ \boolean{showrevisions} } 
        {
            \fcolorbox{red}{white}{#1}
        }
        {
        }
}
\newcommand{\secref}[1]{\hyperref[#1]{Section~\ref*{#1}}}
\newcommand{\figref}[1]{\hyperref[#1]{Figure~\ref*{#1}}}
\newcommand{\tabref}[1]{\hyperref[#1]{Table~\ref*{#1}}}
\rowcolors{2}{gray!10}{white}

\begin{document}


\title[From One World to Another]{From One World to Another: Interfaces for\\Efficiently Transitioning Between Virtual Environments}

\author{Matt Gottsacker}
\affiliation{%
  \institution{University of Central Florida}
  \city{Orlando}
  \state{Florida}
  \country{USA}
}
\email{mattg@ucf.edu}
\orcid{0000-0002-3575-1133}

\author{Yahya Hmaiti}
\affiliation{%
  \institution{University of Central Florida}
  \city{Orlando}
  \state{Florida}
  \country{USA}
}
\email{Yohan.Hmaiti@ucf.edu}
\orcid{0000-0003-1052-1152}

\author{Mykola Maslych}
\affiliation{%
  \institution{University of Central Florida}
  \city{Orlando}
  \state{Florida}
  \country{USA}
}
\email{Mykola.Maslych@ucf.edu}
\orcid{0000-0001-7037-3513}

\author{Hiroshi Furuya}
\affiliation{%
  \institution{University of Central Florida}
  \city{Orlando}
  \state{Florida}
  \country{USA}
}
\email{Hiroshi.Furuya@ucf.edu}
\orcid{0000-0001-7320-5769}

\author{Jasmine Joyce DeGuzman}
\affiliation{%
  \institution{University of Central Florida}
  \city{Orlando}
  \state{Florida}
  \country{USA}
}
\email{jasdeg@ucf.edu}
\orcid{0009-0007-5962-145X}

\author{Gerd Bruder}
\affiliation{%
  \institution{University of Central Florida}
  \city{Orlando}
  \state{Florida}
  \country{USA}
}
\email{bruder@ucf.edu}
\orcid{0000-0003-1450-1802}

\author{Gregory F. Welch}
\affiliation{%
  \institution{University of Central Florida}
  \city{Orlando}
  \state{Florida}
  \country{USA}
}
\email{welch@ucf.edu}
\orcid{0000-0002-8243-646X}

\author{Joseph J. LaViola Jr.}
\affiliation{%
  \institution{University of Central Florida}
  \city{Orlando}
  \state{Florida}
  \country{USA}
}
\email{jlaviola@ucf.edu}
\orcid{0000-0003-1186-4130}

\renewcommand{\shortauthors}{Gottsacker et al.}

\begin{abstract}
Personal computers and handheld devices provide keyboard shortcuts and swipe gestures to enable users to efficiently switch between applications, whereas today's virtual reality (VR) systems do not. 
In this work, we present an exploratory study on user interface aspects to support efficient switching between worlds in VR.
We created eight interfaces that afford previewing and selecting from the available virtual worlds, including methods using portals and worlds-in-miniature (WiMs). To evaluate these methods, we conducted a controlled within-subjects empirical experiment (N=22) where participants frequently transitioned between six different environments to complete an object collection task. 
Our quantitative and qualitative results show that WiMs supported rapid acquisition of high-level spatial information while searching and were deemed most efficient by participants while portals provided fast pre-orientation.
Finally, we present insights into the applicability, usability, and effectiveness of the VR world switching methods we explored, and provide recommendations for their application and future context/world switching techniques and interfaces.
\end{abstract}





\maketitle

\section{Introduction}
\label{sec:intro}

It is common for today's computer users to multitask, or rapidly switch between different applications, e.g., with Alt$+$Tab or \hbox{Cmd$+$Tab} on a personal computer and swipe gestures on a smartphone.
For instance, information workers switch between eight or more application windows~\cite{hutchings2004display} potentially hundreds of times every hour~\cite{olivier2006swish, leijten2014writing} to switch between and complete their tasks~\cite{jahanlou2023task, aral2012information}.
Virtual reality (VR) has shown substantial benefits for many standard computing tasks such as data analysis~\cite{lisle2020thinkspaces}, visualization~\cite{lee2021visceralization}, presentations~\cite{gottsacker2025presentations} and meetings~\cite{lee2025transcriptmeetings} (to name a few).
While modern VR systems allow users to have multiple VR applications on the device, the interfaces for switching between them are far from providing the smooth and continuous mechanisms we enjoy with today's personal computers and smartphones: VR app switching interfaces
require the cumbersome process of navigating 2D menus with pointing, clicking, etc., and do not account for the spatial needs of VR.
VR applications are \textit{entire worlds} with content that can be placed anywhere around the user --- unlike desktop applications, which are constrained to the physical location of a 2D monitor. This unbounded nature of VR arrangement possibilities increases task resumption complexity, especially when the user has to orient themselves before resuming their task, potentially leading to higher cognitive load~\cite{makransky_immersive_2021}.
Additionally, as VR systems are used for more purposes and for longer durations, it is reasonable to assume that in some scenarios, users will expect efficient VR world switching features that provide an experience comparable to the switching features on desktops and smartphones.


In this vein, recent research has explored how multitasking interfaces and user processes can apply in VR, e.g., through the idea of Simultaneous Presence~\cite{ablettSimultaneousPresenceContinuum2025} which explores how users can distribute their attention across multiple virtual environments (VEs) at the same time, or conceptual frameworks designed specifically for quickly switching between VEs~\cite{gottsacker2025xrfirst}.
Additionally, prior work examined the individual elements of the switching process, such as providing previews of user's real-world surroundings~\cite{wang2022lens, budhiraja2015s, seo2024gradualreality} or other environments~\cite{ablett2023point, ablettSimultaneousPresenceContinuum2025, schjerlund2022ovrlap, elvins_worldlets_1998, men_impact_2017, husung2019portals, nam2019wedges}, supporting interaction techniques for selection~\cite{das2024fingerworn, husung2019portals, men_impact_2017, nam2019wedges}, or designing transition effects~\cite{men_impact_2017, husung2019portals, pointecker2022bridging, pointecker2024replica, feld2024simple}.
Yet, to our knowledge, no prior work proposed or evaluated a dedicated interface that unifies these elements into a complete world-switching workflow.
\textbf{In an effort to inform future interfaces for real-world VR multitasking workflows (e.g., switches involving shifts in user context, task, and goal), we present an initial world switching interaction model (see~\figref{fig:interactionmodel}) and derive interfaces that support a continuous workflow of activation, selection of available world options, and transition confirmation.}
In this first structured exploration, we approximate VR app switching as switching between distinct worlds represented as VEs. We used an abstract object search-and-retrieve task across several VEs to compare how UI aspects affect performance and user experience.

With the goal of supporting a continuous interaction across all phases of our world switching model as well as providing a preview, we designed eight world switching interfaces based on this model.
These interfaces vary across preview pattern---World-in-Miniature (WiM) vs. portal-based views---and interaction techniques---Head-based Palette, Hand-based Palette, World Wheel, and Gallery. 
We evaluated these interfaces using a controlled within-subjects experiment with 22 participants who were tasked with rapidly transitioning between different VEs to complete an object retrieval task.
The study task focused on the user interaction with the interface, so the experimental VEs were structurally similar and the search space was well-defined and consistent.
Our empirical investigation includes quantitative measures of efficiency and accuracy, complemented by subjective assessments of usability, cognitive load, user experience, and perceived continuity of interactions.
We found that for our search task, WiMs supported rapid acquisition of high-level spatial information and were deemed most efficient by participants for our task while portals provided fast pre-orientation.
Additionally, our Hand Palette afforded the fastest return to the user's start location.
When combined with the WiM preview, the Hand Palette and World Wheel scored highly for user experience metrics and were preferred by participants.
Our empirical results provide insights into the efficacy of different factors of VR transition interfaces, contribute to the understanding of multitasking in VEs, and pave the path for future development of intuitive, efficient, and fluid world-switching interfaces so that UI solutions are ready when hardware and OS-level support for app switching becomes available. We list our contributions as follows: 
\begin{enumerate}
    \item The first structured exploration (to our knowledge) of interfaces for world switching in VR.
    \item Empirical results from evaluating our techniques using both objective and subjective metrics.
    \item Practical recommendations for researchers and developers on adopting our techniques and creating new ones.
    \item Open-source access to our experimental setup and world switching interfaces.
\end{enumerate}

\section{Background}


VR systems and applications are typically designed around the idea of immersing the user in a single environment at a time. Applications often aim to maximize the sense of presence in that one world, minimizing distractions or links to alternative contexts.
However, as we outline in~\secref{sec:use-cases} below, many scenarios highlighted in the research literature require users to shift their attention and sense of presence across multiple worlds. These shifts can occur frequently, making the design of efficient switching techniques an important concern.
As we show in~\secref{sec:worldswitchrelatedwork}, prior research has begun to explore how to 
make switching smooth and comprehensible, though relatively little attention has been paid to optimizing the efficiency of the switching interaction itself.

\subsection{World Switching Use Cases}
\label{sec:use-cases}

World switching interfaces serve multiple purposes across domains, each presenting unique requirements and interaction patterns. Understanding these use cases provides essential context for designing effective transition mechanisms between virtual worlds. The example use case categories presented in \tabref{tab:usecasestabular} illustrate the breadth of scenarios where world switching occurs, though we do not claim these categories provide the comprehensive coverage of a complete framework.
These use cases and activities demonstrate that world switching is not merely a convenience feature but a fundamental requirement for complex VR workflows. The varying demands across these scenarios emphasize the need for flexible and efficient world switching interfaces that can adapt to different interaction contexts and user requirements.

\setlength{\tabcolsep}{4pt}
\begin{table*}[htp]
\caption{Use case categories and example scenarios that may benefit from frequent world switching in immersive VR applications.}
\label{tab:usecasestabular}
\centering
\footnotesize
\fcolorbox{blue}{white}{
\begin{tabular}{p{0.17\linewidth} p{0.79\linewidth}}
\toprule
\textbf{Category} & \textbf{Examples} \\ 
\midrule
\rowcolor[HTML]{FFFFFF}
\begin{itemize}[leftmargin=1.5em, label={\textbf{(a)}}, labelsep=0.3em, itemsep=0pt, topsep=0pt]
    \item \textbf{Immersed Development} \newline \textbf{\& Content Creation}
\end{itemize}
& 
\textbf{\textit{Environment modeling}} involving sculpting or modifying 3D object positioning requires developers to evaluate their changes from the end-user's perspective, necessitating rapid transitions between editing and preview/test modes~\cite{VRDoh2025, crossrealityDesignInVR2022}. 
\newline\textbf{\textit{Behavior scripting workflows}} require developers to script interactions or adjust physics parameters before switching to test these changes from the user's viewpoint~\cite{zhang2020flowmatic, segura2020vrlearnprog}. 
\newline\textbf{\textit{Record and edit timeline activities}}, where users capture regular or 360 degree videos from multiple virtual world viewpoints, necessitate transitions to timeline edit mode for sound editing and rearrangement, which may require frequently switching 
back to the environment to re-record specific segments~\cite{nguyen2017vremiere, chidambaram2022editAR, griffin20216dive}. \\ \hline

\rowcolor[HTML]{EEEEEE}
\begin{itemize}[leftmargin=1.5em, label={\textbf{(b)}}, labelsep=0.3em, itemsep=0pt, topsep=0pt]
    \item \textbf{Content Streaming}
\end{itemize}
& 
\textbf{\textit{VR Game streaming}} presents complex challenges, as users stream immersive VR experiences while maintaining communication with their audience through minimized chat overlays attached to different anchors like hands or scene objects, with the ability to switch to fuller chat views containing additional audience interaction tools~\cite{yang2025chatgrab, hu2025understanding, wu2023interactions}. 
\newline\textbf{\textit{Prototype demos}} follow similar patterns, where users stream prototype immersive experiences while communicating with stakeholders via voice, requiring smooth transitions between the demo VE and communication applications~\cite{crossrealityDesignInVR2022}. \\ \hline

\rowcolor[HTML]{FFFFFF}
\begin{itemize}[leftmargin=1.5em, label={\textbf{(c)}}, labelsep=0.3em, itemsep=0pt, topsep=0pt]
    \item \textbf{Social Context} \newline \textbf{Switching}
\end{itemize}
& 
\textbf{\textit{Collaborative spaces with multiple people working together}} frequently require users to transition from shared virtual worlds to private virtual spaces for one-on-one conversations, such as when receiving personal meeting invites from a separate application while collaboratively developing immersive experiences in another~\cite{freeman2022working, privateSocialVRConversations2025}. 
\newline\textbf{\textit{Social context breaks}} during brainstorming involves users switching between virtual meetings and personal immersive spaces for individual reflection before returning to the shared environment~\cite{surveyPrivacySocialVR2025, privateSocialVRConversations2025, lisle2020thinkspaces}. \\ \hline

\rowcolor[HTML]{EEEEEE}
\begin{itemize}[leftmargin=1.5em, label={\textbf{(d)}}, labelsep=0.3em, itemsep=0pt, topsep=0pt]
    \item \textbf{Remote Control}
\end{itemize}
& 
\textit{\textbf{Overseeing multiple machines or robots}} in industrial manufacturing chains or smart home systems demands specialized world switching. These scenarios require swift transitions between viewpoints and embodied control modes, allowing users to switch perspectives to take control when needed~\cite{whitney2018ros, ghimire2025avataroid}.\\ \hline

\rowcolor[HTML]{FFFFFF}
\begin{itemize}[leftmargin=1.5em, label={\textbf{(e)}}, labelsep=0.3em, itemsep=0pt, topsep=0pt]
    \item \textbf{Real-world} \newline \textbf{Place Viewing}
\end{itemize}
& 
\textbf{\textit{Urban planning}} workflows involve switching between multiple view modes and data visualization layers of a city plan to gather comprehensive information for construction decisions~\cite{zhang2021urbanvr, yang2021tilt, spur2020mapstack}. 
\newline\textbf{\textit{Location previewing}} for travel planning requires users to transition between multiple real world place previews and nearby attractions to make comparative assessments for trip planning~\cite{kieanwatana2024virtual, ye2024investigating}.\\ \hline

\rowcolor[HTML]{EEEEEE}
\begin{itemize}[leftmargin=1.5em, label={\textbf{(f)}}, labelsep=0.3em, itemsep=0pt, topsep=0pt]
    \item \textbf{Cross-Application} \newline \textbf{Interoperability}
\end{itemize}
 & 
\textbf{\textit{Environment object reuse workflows}} require users to maintain multiple 3D scenes simultaneously, necessitating frequent world transitions, for example, when copying objects from one scene and pasting them into another~\cite{vrVersionControl2023, cools2022blending}. \newline
\textbf{\textit{Issue diagnosis processes}} often demand switching between multiple applications to troubleshoot system malfunctions~\cite{liu2023challenges}, such as avatar rendering problems across different applications.\\ \hline
\end{tabular}
}
\Description{Table listing seven categories of VR use cases that may benefit from frequent world switching, each with example scenarios. (a) Immersed Development and Content Creation includes tasks like environment modeling, behavior scripting, and record or edit timeline activities that require developers to view changes from the user’s perspective. (b) Content Streaming includes VR game streaming and prototype demos where transitions between live and communication applications occur. (c) Social Context Switching includes collaborative spaces and social context breaks where users transition between shared and private virtual spaces. (d) Remote Control covers overseeing multiple machines or robots and switching viewpoints to take control when needed. (e) Real-world Place Viewing includes urban planning and location previews requiring multiple world transitions. (f) Cross-Application Interoperability includes environment object reuse workflows and issue diagnostics requiring switching between different applications. Each category provides representative tasks illustrating how world switching supports efficient transitions across virtual contexts.}
\end{table*}

From these use cases we focus on a few common features:
\textit{1) Previews and Provision of Spatial Context. }
We observe that some workflows demand detailed spatial previews to assess 3D object positioning (a) while others prioritize contextual information about the nature of the virtual environment and who is in it (c). 
In other use cases, users need simultaneous visibility into the states of multiple environments (d, e), making it clear that world switching should support efficient provision of the spatial context for the task.
\textit{2) Interaction.}
The switches are frequent (a, f), smooth (b), and swift (d). 
It follows that the interactions used to trigger these should be intuitive and simple to use.
\textit{3) Transition.}
Inherent in switching between worlds is the moment where one world becomes the next. 
In demanding tasks (a, f) these transitions are near-instantaneous to minimize workflow interruption, while in social applications (c) users may find it desirable to take more time. \textit{4) Return to home.}
Many use cases involve several different views or environments (a, c, d, e, f). 
It is usually the case that there is a primary environment around which the world switches revolve. 
For example, in a development context (a) there may be a primary editor environment that users return to often, or in robot supervision (d) there may be a single environment serving as the overview dashboard. Hence, the world switching interface must support quick formation of spatial memory~\cite{scarr2013spatialmemory} to perform a return-to-home action.
To identify strategies for implementing solutions to these needs, we examined the literature for works that address them in world switching and adjacent domains.


\subsection{World Switching Interaction Techniques and Transition Effects}
\label{sec:worldswitchrelatedwork}


Recent work touches upon three elements of world switching: previews, interaction, and transition, but rarely treats them as a single, continuous workflow. Here we synthesize adjacent findings to motivate our design choices and study.

\subsubsection{Previews: Additional Spatial Context}

Previews help provide additional spatial context to the user about a destination before switching.
A classic approach for spatial previews is commonly achieved with portals (additional windowed viewpoints).
In 1998, Elvins et al.~\cite{elvins_worldlets_1998} used ``worldlets,'' or interactive 3D thumbnails in a 2D web browser context, to provide users with multiple 3D views of a virtual city to assist them in a wayfinding task. 
Although these viewpoints are of a single virtual world, the idea of using portal-like limited views to preview other worlds remains significant and is well-represented in current literature, including works by Men et al.~\cite{men_impact_2017} and Husung and Langbehn~\cite{husung2019portals}.
Recently, Linne et al.~\cite{linne2025slivr} introduced SliVR, a hub that arranges several portal previews of different VEs around the user for multitasking purposes.
Taking a different approach to spatial previews, Schjerlund et. al.~\cite{schjerlund2022ovrlap} proposed the OVRLap technique for overlaying multiple semitransparent viewpoints from different locations in a VE, with only the selected viewpoint opaque and interactable by the user.
Combining this concept with portals, Ablett et al.~\cite{ablettSimultaneousPresenceContinuum2025} introduced techniques that allowed a user to preview a second VE in full visual detail through a portal while overlays that extended beyond the portal bounds provided partial visual detail of the second VE (e.g., through reduced opacity or stencil effects).
Like other examples of portal-based previews, these techniques implicitly involve \textit{interaction} by the user to activate the portal, reposition it in space to preview particular parts of the VE, and finally deactivate it.

Another classic technique for providing additional spatial context is the world-in-miniature (WiM)~\cite{stoakley1995virtual}, where the user can see and interact with a miniature model of a VE to rapidly inspect it from multiple viewpoints.
Nam et al.~\cite{nam2019wedges} used WiMs, in addition to portals, to help users understand which world they are selecting and simultaneously view datasets from different forests.
Techniques combining WiM and portal previews, particularly for traveling between disparate locations in a large VE, were shown by Elvezio et al.~\cite{elvezio2017preorient}. 
In their paper, participants used a WiM to select a location within the large-scale environment to teleport to, then used a portal to preview the post-teleportation view and fine-tune the post-teleportation location and orientation. In our work, we expect that users in task-driven contexts will need to not only switch between virtual worlds, but also preview and preorient themselves before transitioning. Hence, we include both WiM and portal metaphors in our exploration of world switching techniques.

\subsubsection{Interactions for Selection of World Targets}
We review selected interaction techniques related to world switching concepts.
Das et al.~\cite{das2024fingerworn} explored finger-based techniques for traveling up and down Milgram's Reality-Virtuality Continuum~\cite{milgramTaxonomyMixedReality1994}.
In their study, participants wore a set of buttons, each mapped to switch the user to a view of the real world, augmented reality, augmented virtuality, and VR. 
In Husung and Langbehn's work~\cite{husung2019portals}, users could trigger transitions between two VEs by walking through a portal or grabbing and pulling an orb-shaped portal towards themselves. 
Men et al.~\cite{men_impact_2017} also used locomotion through portal to initiate a transition between two virtual worlds.
In Nam et al.'s WiM implementation~\cite{nam2019wedges}, users grabbed and dragged WiMs to re-order portal views.
Considering world switching interactions for contexts described in Section~\ref{sec:use-cases}, we focus on traditional 3D interaction techniques, such as grabbing and pulling, rather than locomotion or the use of additional wearables.

\subsubsection{Transition Effects}
Methods for transitioning a user's original viewpoint to a destination viewpoint include effects seen in film and videos, such as cuts, fades, and swirls. 
Men et al.~\cite{men_impact_2017} evaluated the effects of such visual transitions on presence as users walked through a portal that teleported them between two VEs.
Husung and Langbehn~\cite{husung2019portals} expanded this work with more complex transitions, including a \textit{transformation transition}, where a small oval-shaped ``rift'' revealing the destination world grows until it completely covers the user's field of view, a \textit{portal transition}, where the user walks directly into the new environment, and an \textit{orb transition}, where an orb that displays the new VE wraps around the user's head.
Pointecker et al.~\cite{pointecker2022bridging, pointecker2024replica} found that while users praised the excitement of a portal technique, for frequent reality transitions they preferred a simple fade, finding it most suitable for ``conventional applications.''
Feld et al.~\cite{feld2024simple} corroborated these findings for continuously transitioning between two VEs while performing a demanding memory game task, where participants preferred the efficiency of fast, abrupt transitions, at the cost of reduced presence and continuity.
Given the established preference for efficiency in frequent, task-driven world switching contexts that we explore, we use the fade transition~\cite{feld2024simple, feld2023keep}.

\begin{figure*}[t]
    \centering
    \includegraphics[width=0.7\linewidth]{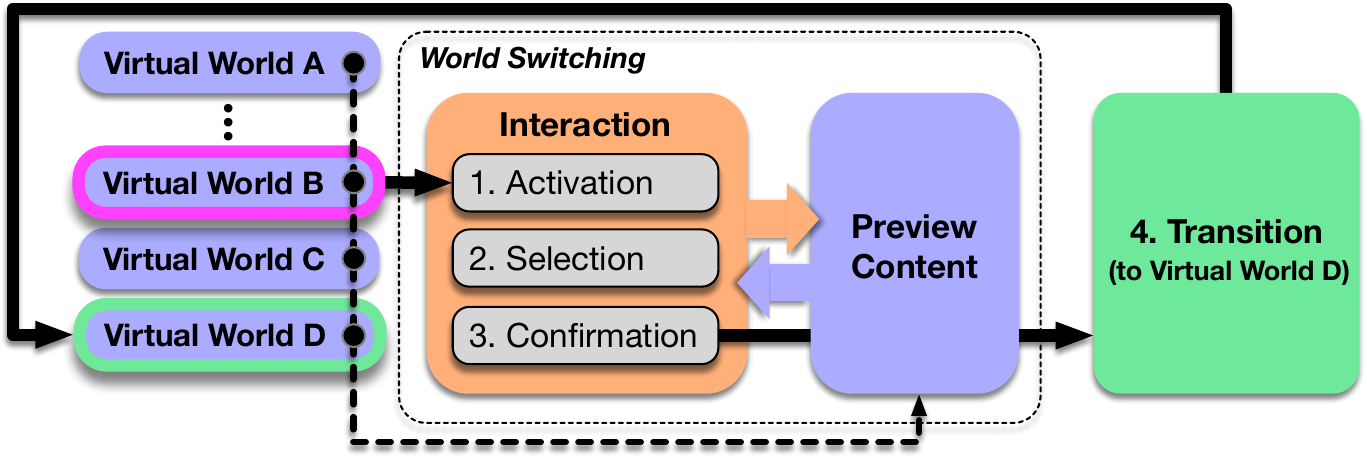}
    \caption{Interaction model for switching worlds. The user begins in Virtual World B (highlighted in magenta) and initiates the switching process. World switching proceeds through an interaction sequence with three steps: \textbf{(1) Activation} of the switching interface, \textbf{(2) Selection} of a target world from the available options, and \textbf{(3) Confirmation} of that choice. During these steps, the system may provide preview content to help the user evaluate the destination. Once confirmed, the user begins the \textbf{(4) Transition} to the selected target world (e.g., Virtual World D). This model distinguishes user-driven interaction steps from system-provided previews and transitions.}
    \label{fig:interactionmodel}
    \Description{Flow diagram of the world switching process across multiple virtual environments. On the left, four virtual worlds are listed vertically (A, B, C, and D). Arrows show that a user begins in one world (highlighted world B) and initiates world switching. The central box is labeled “World Switching” and contains an interaction sequence of three ordered steps: (1) Activation, (2) Selection, and (3) Confirmation. These steps feed into a “Preview Content” element. A final arrow connects this preview to step (4) Transition, shown as moving the user into the target virtual world (world D). Solid and dashed arrows indicate flow of actions and return paths between worlds and the switching process.}
\end{figure*}

\vspace{-1ex}
\subsection{Research Gap}
Despite the diversity of transition and interaction techniques represented in the literature, a gap remains in the study of world-switching interfaces on their own, especially with regard to their effectiveness in supporting frequent task-based switches between disparate immersive contexts.
In desktop and mobile devices, the experience of switching between applications and task contexts consists of preview, interaction, and transition, each contributing to a sense of efficiency and ease of use. In this work, we directly address the interface of switching worlds, first by offering an interaction model for designing these interfaces, then prototyping several using this model, and finally evaluating them in a user study using both objective and subjective metrics. 

\section{Creating our VR World Switching Interfaces}

Our world switching interface design approach was informed by task-switching paradigms in desktop computing.
Building on these inspirations, we constructed an interaction model that characterizes the essential components and processes of switching between VR worlds.
Along with the takeaways from our use cases (\secref{sec:use-cases}) and gaps in related world switching interface designs (\secref{sec:worldswitchrelatedwork}), this model provided a foundation for articulating the motivating design questions that guided our exploration (\secref{sec:designQuestionsandInteractionModel}).
Finally, we implemented selected prototypes to enable systematic evaluation through a controlled user study.

\subsection{Drawing Inspiration From Desktop for VR World Switching}
\label{sec:research-questions}

Initially, desktop task switching caused cognitive overhead~\cite{10.1145/800045.801580}, which was resolved as interfaces matured over time. Fast task switching, task resumption and ease of re-acquiring mental task context were pointed out as crucial requirements for efficient task switching~\cite{card1986switching}. Modern desktop interfaces fulfill these and offer efficient task switching through shortcuts like Alt+Tab/Cmd+Tab and three-finger swipes\footnote{e.g., Alt+Tab / Cmd+Tab for Task Switcher / App Switcher, three-finger touchpad swipe for Task View / Mission Control}. They make use of a \preview{} to allow users to see the gist of an application before selecting it, and incorporate easy to learn and efficient \interaction{} to cycle through and select a certain application.
Importantly, the activation of the switching interface, the selection of the application to switch to, and the confirmation are all part of a \textit{continuous interaction flow}.
For example, on Windows, when the user holds down the Alt key and presses the Tab key, a wraparound list/grid with a live \preview{} of each application in the user's current workspace appears, and the first application in the list is outlined to indicate it is selected.
While still holding down the Alt key, pressing Tab navigates to the next app in the list.
Releasing the Alt key confirms the switch: the selected app is brought to the foreground and focused.
While these mechanisms are deeply embedded in everyday practice, their underlying principles (e.g., rapid access, context recovery) remain underexplored in immersive environments.

\subsection{VR World Switching Interaction Model and Design Questions}
\label{sec:designQuestionsandInteractionModel}

As mentioned in~\secref{sec:intro}, the spatial arrangement complexity of VR worlds introduces unique challenges for achieving a smooth and continuous switching experience.
Thus, reaching efficient world switching in VR requires prototyping and testing various VR world switching interfaces.
Inspired by the efficiency of desktop task switching, we introduce the \textbf{interaction model} shown in~\figref{fig:interactionmodel} to break down the primary steps and components of VR world switching interfaces.
Here, world switching begins with the \activation{} of the world switching interface, which displays the \previews{} of the worlds the user may visit.
Then, the user \selects{} between candidate worlds
and \interacts{} with the world \previews{} until the user is ready to \confirm{} their choice and begin the \transition{} to the desired world.
Each of these steps involves numerous UI design choices accounting for broader goals (e.g., balancing user experience with implementation complexity). However, unlike the \transition{} aspect of VR world switching, the \interaction{}, \preview{}, return-to-home (see~\secref{sec:use-cases}) and pre-orientation aspects have not received much attention in literature. Hence, our motivating design questions are as follows: 
\begin{enumerate}
    \item Which world \preview{} visualizations support efficient visual search and pre-orientation?
    \item Which \interaction{} techniques leveraging different reference frames support efficient visual search and return-to-home?
\end{enumerate}

\subsection{World Switching Interface-design}
\label{sec:derived-techniques}
Guided by our motivating questions, our approach for designing interfaces focused on frequent world switching rather than any particular task the user performs between transitions. 
Our methodology incorporates 
eight
interfaces by combining two \textbf{preview patterns} (\textit{portal} and \textit{WiM}) with 
four
\textbf{interaction techniques} described in the subsections that follow (\textit{Hand-based Palette}, \textit{Head-based Palette}, \textit{Gallery}, and \textit{World Wheel}).
All interfaces included a centrally located preview and an array of selectable environment options as static thumbnail images with a cyan background (current environment thumbnail was non-selectable and grayed-out).
All interactions were performed exclusively via hand tracking.
The transition effect between worlds was consistent across all interfaces. Once users confirmed a world to switch to, their view faded to black and then faded in to the selected world, with each fade lasting $0.5s$. This transition was chosen to be in accordance with related work that suggests using a short fading transition for task-based contexts~\cite{feld2024simple}.
A video demonstration of the interfaces is available in the supplementary materials and the interfaces, experiment setup, and anonymized collected metrics are available via
\underline{\url{https://github.com/mott-lab/WorldSwitchUI}}.

\subsubsection{\textbf{Preview Pattern}}
\label{previewPattern}
To explore how to best visualize a world to a user before they transition there, 
we employed two 3D preview patterns prevalent in the research literature: \textbf{\textit{portal}} and \textbf{\textit{world-in-miniature (WiM)}}.
These previews are commonly used to provide spatial information about an environment or location where a user is not currently located
to provide affordances such as distant object interaction~\cite{ablett2023point, danylukDesignSpaceExploration2021, stoakley1995virtual, lisle2022clean}, collaboration~\cite{wang2025teamportal}, navigation~\cite{danylukDesignSpaceExploration2021, mine1995virtual}, visualizing complex spatial relationships~\cite{chittaro2005breakaway}, and balancing awareness or presence across multiple locations~\cite{nam2019wedges, ablettSimultaneousPresenceContinuum2025, schjerlund2022ovrlap, linne2025slivr}.
\begin{itemize}[leftmargin=1em, labelsep=0.3em, itemsep=2pt, topsep=0pt]
    \item \textbf{Portal}: The portal preview provides a first-person window into the selected environment at full-scale~\cite{husung2019portals, bruderportal, deering1993making, laviola20173d}.
    Our portal continuously updates the view of that environment from the user’s perspective and renders only the content of the previewed environment (in stereo 3D) by rendering the graphics layer assigned to that environment to a circular preview frame.
    \item \textbf{WiM}: The WiM provides a 3D scaled-down representation of a specified environment~\cite{stoakley1995virtual,laviola20173d, jerald2015vr}.
    Our WiMs of the world options were scaled to $1/100$ of each world's actual size and occupied a circular area with a radius of $0.11$ meters. 
    Our WiM previews also included an abstract representation of the user as a purple avatar with a head, torso, and arms.
    The local positions of the avatar's head and arms were updated with the user's own head and hands.
\end{itemize}

\noindent Both preview patterns provide advantages for visual search by allowing the user to view aspects of a world without requiring them to fully transition there.
The pre-orientation affordances of both of these preview patterns depend on the user's perspective and position with respect to the world being previewed.
That is, when looking through a portal preview of another world, the user can see full-scale, first-person view of everything they will be facing immediately after they transition.
When looking at a WiM preview, the third-person preview allows the user to obtain a more global perspective and see objects that will surround them after they transition to the previewed world.
The tracked avatar allows the user to see how they would be oriented with respect to the world before transitioning there~\cite{elvezio2017preorient}.

\begin{figure*}
  \centering
  \includegraphics[width=\linewidth]{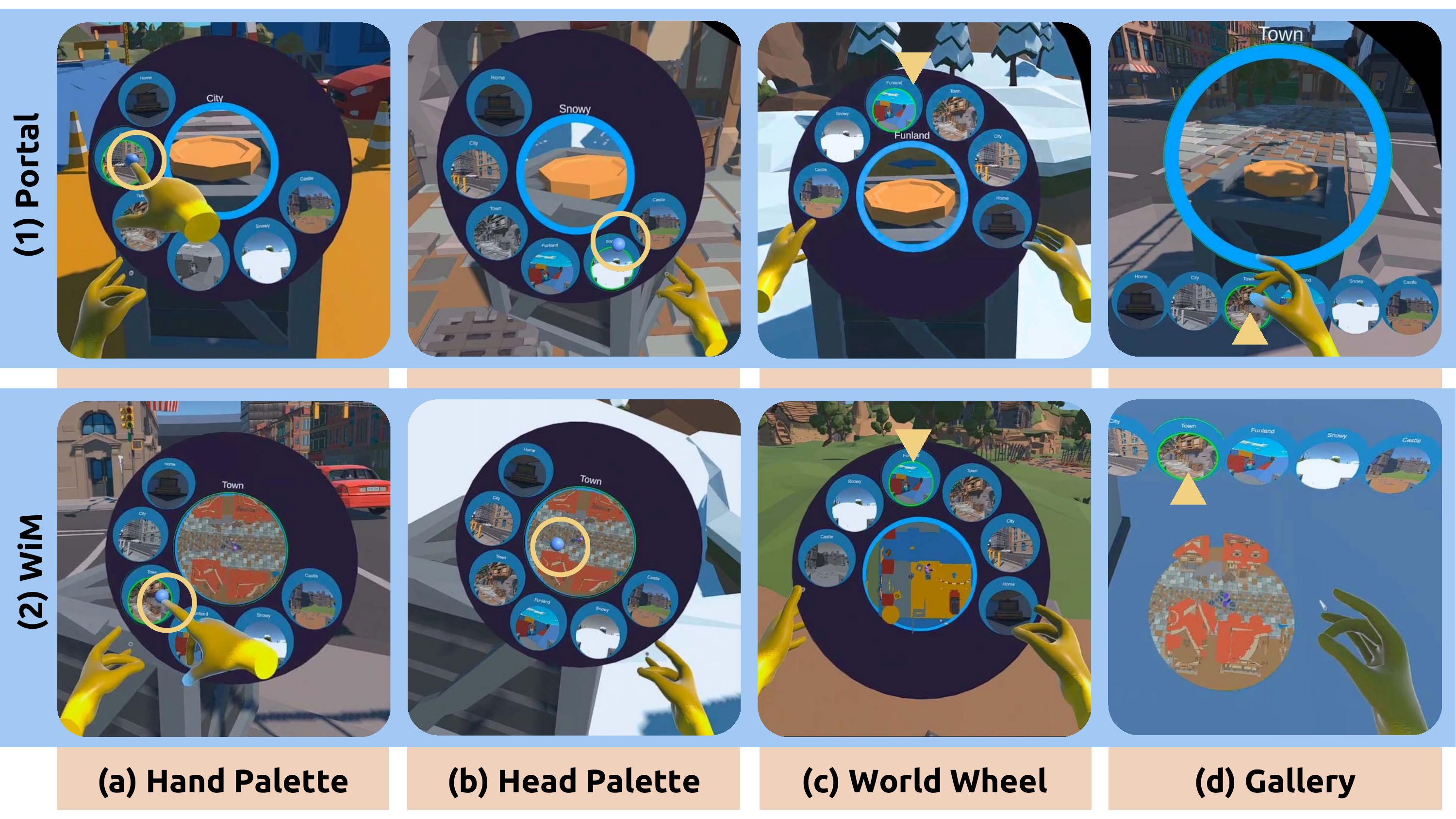}
  \vspace{-5ex}
  \caption{Eight virtual world switching interfaces evaluated in our study.
  Two preview patterns are used: (1) Portal and (2) World-in-Miniature (WiM). Four interaction techniques were derived: (a) Gallery, (b) Head Palette, (c) Hand Palette, and (d) World Wheel. The selected environment is indicated by a yellow arrow---(a) and (d)---or a blue spherical cursor---(b) and (c). The latter is highlighted with a yellow circle for clarity.}
  \label{fig:teaser}
  \Description{Screenshots of six world switching interface techniques in virtual reality. The figure is arranged in two rows by interaction metaphor: Portal (top row) and World-in-Miniature (bottom row). Each column shows a different interface layout: (a) Gallery, (b) Head Palette, (c) Hand Palette, and (d) World Wheel. In the Portal row, each interface appears within or adjacent to a large circular portal that frames the next virtual world. A circular menu of candidate worlds surrounds the portal, with the user’s hand visible pointing or selecting among them. In the World-in-Miniature row, miniature world previews are displayed in front of the user’s hands or body. The Gallery shows small floating spheres representing worlds. The Head Palette positions the circular menu near the user’s head. The Hand Palette places the same menu near the user’s hand. The World Wheel shows the menu floating in front of the user, with the currently selected world enlarged at the center. Each variation highlights different ways of previewing and selecting among available worlds before switching.}
\end{figure*}

\subsubsection{\textbf{Interaction Technique}}
\label{sec:interactionMetaphor}

To explore how to best support efficient search and return-to-home interactions, we aimed to support both primary components of our world switch interaction model (world selection and preview interaction) in a single, continuous user interaction flow with a spatially consistent layout~\cite{czerwinski1999thumbnail, robertson1998data}. The full world selection process involves activating the world switch UI, selecting between different world options, and confirming the world option to transition to. 
The other primary component of the world switch interaction involves using a preview for visual search and pre-orientation.
For the preview interactions, we focused on efficiency in moving the portal to see different views or understanding the spatial relationship of the user's miniaturized avatar to the WiM content.

As these preview interactions are inherently embodied, we decided to focus our interaction design on body-based \textit{reference frames} (the spatial coordinate systems relative to which the interface’s virtual content and interactions are defined).
To keep this first exploration focused, we considered three user-centered reference frames common in 3DUI interaction techniques: one hand, both hands, and head~\cite{laviola20173d, maslych2024research,jerald2015vr}. 
Based on our interaction model that separated the interface preview and interaction, we enumerated all possible combinations that varied each of these reference frames for display and interaction separately.
For each combination shown in \tabref{tab:ref-frame}, we used iterative sketching and prototyping to arrive at interfaces that integrate the preview and world selection patterns into a unified technique with continuous interaction, guided by the need to support visual search and pre-orientation (see \secref{sec:designQuestionsandInteractionModel}).

Rows with an orange background in \tabref{tab:ref-frame} indicate interaction techniques we investigated in the main study and are described in detail below.
Rows marked as Implausible in the Technique column are combinations of preview / interaction reference frames that seemed implausible to create or unintuitive for the user if created, so we did not implement them. 
Rows marked as P1 and P2 are techniques we implemented and piloted but ultimately excluded from the user study because they were similar to our other techniques (see \secref{sec:considerations}).
We also included a baseline technique in the user study inspired by interfaces for switching worlds in contemporary VR headsets (i.e., app menus without live previews). \\

\begin{table}[t]
\vspace{-2ex}
\caption{Combinations for reference frames of display elements and interaction flow in world switching interfaces. Orange backgrounds indicate techniques that we implemented and evaluated in the present user study. Technique columns marked (Implausible) indicate combinations we excluded due to being implausible or requiring non-intuitive or cumbersome interaction.
(P1) and (P2) represent techniques implemented but excluded during pilot testing due to not offering meaningful differences from our other proposed techniques. Our full exclusion rationale are explained in~\secref{sec:interactionMetaphor}.}
\footnotesize
\vspace{-0.5ex}
\begin{tabular}{lll}
\hline
\textbf{Preview Reference Frame} & \textbf{Interaction Reference Frame} & \textbf{Technique} \\ \hline
\rowcolor[HTML]{FFFFFF}
Hand       & Other Hand & \cellcolor[HTML]{eed1bc} \handPalette{} \\ \hline
\rowcolor[HTML]{FFFFFF}
Hand       & Both Hands & \cellcolor[HTML]{EEEEEE} $\times$ (Implausible) \\ \hline
\rowcolor[HTML]{FFFFFF}
Hand       & Head       & \cellcolor[HTML]{eed1bc} \headPalette{} \\ \hline
\rowcolor[HTML]{FFFFFF}
Both Hands & One Hand   & \cellcolor[HTML]{EEEEEE} $\times$ (Implausible) \\ \hline
\rowcolor[HTML]{FFFFFF}
Both Hands & Both Hands & \cellcolor[HTML]{eed1bc} \steeringWheel{} \\ \hline
\rowcolor[HTML]{FFFFFF}
Both Hands & Head       & \cellcolor[HTML]{EEEEEE} $\times$ (P1) \\ \hline
\rowcolor[HTML]{FFFFFF}
Head       & Hand       & \cellcolor[HTML]{eed1bc} \gallery{} \\ \hline
\rowcolor[HTML]{FFFFFF}
Head       & Both Hands & \cellcolor[HTML]{EEEEEE} $\times$ (P2) \\ \hline
\rowcolor[HTML]{FFFFFF}
Head       & Head       & \cellcolor[HTML]{EEEEEE} $\times$ (Implausible) \\ \hline
\end{tabular}
\vspace{-1.3ex}
\label{tab:ref-frame}
\Description{Table showing combinations of Preview Reference Frame, Interaction Reference Frame, and the resulting Technique. The table lists eight possible frame pairings. When the preview is attached to the hand and interaction is performed with the other hand, the technique is Hand Palette. When the preview is hand-based and interaction uses the head, the technique is Head Palette. When both preview and interaction use both hands, the technique is World Wheel. When preview is attached to the head and interaction uses one or both hands, the techniques are labeled Gallery and P2 respectively. Other combinations are marked as implausible. The table illustrates which frame pairings yield viable world switching techniques.}
\vspace{-2ex}
\end{table}

\noindent
\textbf{(T1) Hand-based Palette}

\textbf{\textit{Content:}}
The interface is a circular palette that remains anchored to the hand and moves with it. 
The palette has a diameter of $0.5$m and is composed of a central preview circle surrounded by an outer ring of smaller option circles. The central display provides a live preview of the currently selected environment and clearly displays its name at the top.
Each of the surrounding option circles have a diameter of $0.11$m and feature a static image and the corresponding environment name. 

\textbf{\textit{Interaction:}} To enable the interface, the user performs a thumb--middle finger pinch on the left hand.
The palette remains active as long as the left-hand pinch is maintained; releasing the pinch immediately deactivates it. While active, a small blue sphere (diameter $0.025$m) appears at the tip of the right index finger to assist with precise interaction. To preview an environment, the user simply touches the appropriate option circle, which instantly is outlined in green to indicate current selection, and the central preview updates accordingly. The interface is rendered in a dark blue color to clearly differentiate it from user surroundings. Once the user is satisfied with their choice, they trigger the transition by touching the selected environment icon or the center preview circle while simultaneously releasing the left-hand pinch. \\

\noindent
\textbf{(T2) Head-based Palette}

\textbf{\textit{Content:}}
Same as Hand-based Palette. 

\textbf{\textit{Interaction:}} A right-hand thumb-middle finger pinch activates the head-based palette enabling a round palette anchored to the right hand of the user and follows its movements.
The palette is active only when the right hand pinch is maintained; release of the pinch deactivates it immediately. Unlike the hand-based palette that uses a blue sphere on the tip of the right hand index finger for precise interaction, the head-based palette uses a blue sphere projected from the user's head gaze onto the palette and serves as a precise interactor with it. If this projected sphere is over any circle of the ring of environment options, the matching circle of that environment turns green, and identical feedback is provided if the interactor is moved over the central preview circle. The transition is confirmed when the user faces an environment circle (either within the ring or at the center preview of it) and releases the right-hand thumb–middle finger pinch simultaneously, thereby switching to the desired VE. \\


\noindent
\textbf{(T3) World Wheel}

\textbf{\textit{Content}:}
The interface has a dark blue circular background with a ring of environment thumbnails around a central preview.
This wheel is positioned at the midpoint between the user's hands and has a diameter that dynamically extends from one hand to the other.
Each thumbnail is a small circle that represents an environment option and includes a static representative image and the environment name displayed at the top of the circle. 

\textbf{\textit{Interaction:}}
The activation is by a pinch of the right-hand thumb and middle finger. 
Users select an environment by rotating the wheel left or right, aligning the desired thumbnail with a yellow triangular cursor at the wheel’s apex. When a thumbnail collides with the cursor, it is outlined in green and the central preview updates synchronously, with the currently previewed environment name shown above. The wheel’s size
is adjustable by changing the distance between the user’s hands. The preview circle has a diameter of $0.38\%$ of the wheel and each thumbnail circle has a diameter of $0.22\%$ of the wheel. 
To confirm a transition, the user releases the pinch when the desired environment is under the cursor.
If the cursor is not intersecting with any thumbnail (i.e. none is selected) the interface is deactivated. \\

\noindent
\textbf{(T4) Gallery}

\textbf{\textit{Content:}}
The interface is head-referenced and always located at $0.8$m from the head, composed of a central circular portal (diameter $0.6$m) displaying a live preview of the current selected environment with a horizontal array of circular thumbnails positioned below (diameter $0.15$m). Each thumbnail shows a static representative image of an environment and its name above it. 

{\textbf{\textit{Interaction:}}}
The interface is activated by a right-hand thumb--middle finger pinch and remains active while the pinch is maintained.
A yellow triangular cursor slides horizontally in sync with hand movement; the thumbnail under the cursor is highlighted in green, simultaneously updating the central preview. Additionally, a speed gain mechanism allows rapid navigation along the options, with the cursor clamped at the boundaries of the environment option list when further movement in either direction is attempted. 
Releasing the pinch at the same horizontal level as the activation pinch snaps the cursor to the nearest thumbnail and keeps the interface active without triggering the transition (to enable clutching).
Releasing the pinch with an upward slide $0.1$m above the pinch initiation position confirms the transition, while a downward release $0.1$m below cancels it and disables the interface.
While the pinch moves vertically, the cursor remains in place.
Full details of our scrolling implementation, including equations describing non-isomorphic gains, can be found in the Supplementary Materials. \\


\noindent
\textbf{(Baseline) Point-to-Select}
\label{baseline}

\textbf{\textit{Content}:}
The interface consists of two horizontal rows of evenly spaced circular options (3 per row in our context, each $0.15$m in diameter), each representing a different environment.
When enabled, the interface is positioned $0.8$m distance in front of the user's right hand and rotated to face toward the user.
The interface is world-referenced, maintaining the same orientation and position from the moment it is spawned.
Each circle contains an environment thumbnail with the name of the environment positioned above. 

\textbf{\textit{Interaction}:}
The baseline technique activates when the user performs a right-hand thumb-middle finger pinch. The UI remains active as long as the pinch is maintained. 
The user can select an option by pointing at it, and if the pointer lands on a circle, the circle is immediately outlined in green for feedback. A transition to the selected environment occurs when the user releases the pinch while pointing at the desired option.

The design of the technique was inspired by common VR application selection screens that arrange app icons in grid patterns.
We used this as a comparison baseline as ray casting~\cite{baloup2019raycursor, mine1995virtual} is one of the most common techniques for selection in VR~\cite{laviola20173d, jerald2015vr}, and represented a technique without preview used to compare against our portal and WiM preview-based techniques 
(see ~\secref{previewPattern}).

\begin{figure*}[t]
    \centering
    \includegraphics[width=\linewidth]{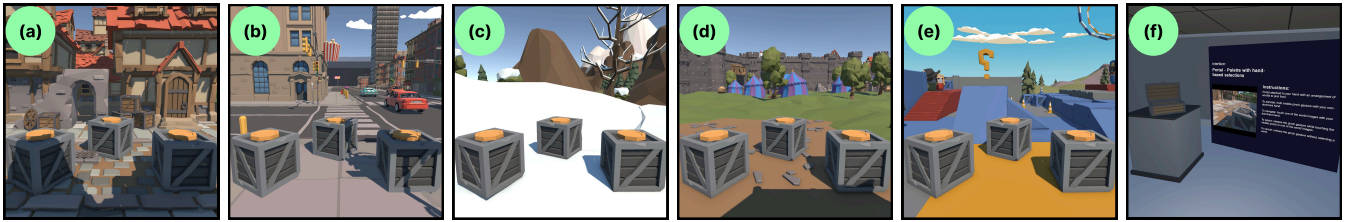}
    \caption{Virtual environments (VEs) used in our experiment. (a-e) VEs where coins could appear, (f) Starting location (Home environment) where participants deposited coins and received interface training before the start of each trial block for a technique through a UI containing text instructions and a video demonstration.}
    \label{fig:environments}
    \Description{Screenshots of six distinct virtual environments used in the study, each labeled with a green number. a) A medieval village street with stone steps, timber-framed houses, and crates in the foreground. b) A modern city street with tall buildings, cars, and crates arranged along the sidewalk. c) A snowy mountain scene with snow-covered trees and crates placed on the ground. d) A medieval fairground with tents and castle walls in the background, with crates placed in the foreground. e) A colorful obstacle-course environment with ramps, rings, and crates positioned on platforms. f) An indoor gallery space with a display pedestal holding a crate and an informational wall panel describing the exhibit. Each environment shows variations in theme and setting while maintaining consistent interactive crate objects.}
\end{figure*}

\subsubsection{\textbf{Pre-study Pilot Testing and Additional Considerations}}
\label{sec:considerations}

We conducted informal pilot testing with ($n = 8$) experts (VR interaction researchers) to fine-tune our techniques (e.g., scrolling speed gains), assess that they were distinct enough from one another, and test their practical utility. We excluded two techniques from the main study. \textbf{(P1) World Wheel with Head Cursor} (\textit{Content}: same as World Wheel; \textit{Interaction}: same as Head-based Palette) was excluded  based on pilot participants reporting it as very similar to the Hand-based Palette technique, except more cumbersome as it required extending both arms instead of just one ($n = 6$). \textbf{(P2) Gallery with World Wheel Selection} (\textit{Content}: preview same as Gallery, world options displayed same as World Wheel; \textit{Interaction}: same as World Wheel) was excluded due to reports that the preview's placement outside of the world wheel made it difficult to see ($n = 7$). Pilot users preferred world options and previews being both displayed relative to the same reference frame (head or hand).

The set of techniques we derived (\tabref{tab:ref-frame}) is not exhaustive of all the possible reference frame combinations and we do not claim that our implementations are the best ways to capture those combinations. For example, other reference frames are worth investigating (e.g., world, torso, tracking space, etc.). All our techniques were adapted for both left‐ and right‐handed users to ensure that any precise interactions relied on the user’s dominant hand, though we describe them from the perspective of a right-handed user. We were interested in hand-tracking-based interactions for this study which previous work has shown to be preferred for direct touch interaction~\cite{luong2023controllershands}, so all interfaces used thumb and middle finger pinch gestures, which are directly mappable to physical hold and release switch actions on physical controllers~\cite{maslych2024research}. Both implementations are provided in the supplemented repository (see \secref{link:open-source}).

\section{Study Methodology}
\label{sec:method}
 

We conducted an exploratory study evaluating our world switching interfaces in a within-subject study ($n=22$) with an object retrieval task (\secref{sec:task}) that required switching between worlds.
We designed multiple VEs in Unity to simulate worlds for our study, and the world switching interface as an interactive object that manages those VEs (see \secref{sec:apparatus}).
Our independent variable was the switching interface.
Our world switching design process (see \secref{sec:derived-techniques}) led to eight interfaces: each of our four interaction techniques implemented for both portal and WiM preview patterns.
We compare our proposed interfaces with a common baseline point-to-select method (see~\secref{baseline}).
We assessed participant completion times and subjective perception (\secref{sec:metrics}).

\subsection{Study Task}
\label{sec:task}


We designed a task to answer our guiding questions (\secref{sec:research-questions}) about world switching interfaces supporting continuous interactions involving \textbf{searching}, \textbf{pre-orienting}, and \textbf{returning-to-home}, inspired by use-cases involving switching between worlds mentioned in \secref{sec:use-cases}.
We adapted an object collection task from the OVRlap evaluation by Schjerlund et al.~\cite{schjerlund2022ovrlap}, originally used to evaluate techniques for perceiving and interacting across multiple distant locations simultaneously. 
The object collection task required participants to quickly find a coin in one of five VEs, grab it, and transition back to the home VE to deposit it, allowing us to break down the completion time metrics into \textbf{search}, \textbf{retrieve} and \textbf{deposit times}. 
To align with our goal, we made grabbing the coin easy by-design, allowing participants to focus on transitions, and reduce the influence of the object selection technique on the retrieval time.

Every task began with the participant centered in the home world (see~\figref{fig:environments} (f)). The trial coins were gold and placed atop one of 
three
crates within one of 
five
distinct VEs (excluding the start home VE). One coin was collected per trial and coin placement varied in each trial to avoid spatial memory repetition, ensuring it never appeared in the same environment or on the same crate consecutively. Coins could be grabbed with either hand using a direct index-thumb pinch. Upon grabbing, coins were outlined in green as visual confirmation and immediately snapped into a nearby, non-obstructive storage unit~\cite{maslych2024research}. This storage unit was in the form of a small sphere anchor that temporarily stored the coin, allowing participants to use their hands freely for manipulating the transition technique. After collecting the coin, participants transitioned back to the home world and deposited the coin into the chest. A blue outline around the chest was triggered when the coin intersected it, and upon successful deposits, a sound effect played.

For our study task, we created six VEs (city park, amusement park, castle, marketplace, and snowy landscape) using Unity Asset Store assets\footnote{\href{https://assetstore.unity.com/packages/3d/environments/polygon-sampler-pack-207048}{Synty Studios POLYGON - Sampler Pack}} (see~\figref{fig:environments}). Three wooden crates, where trial coins spawned, were positioned similarly across all VEs: one in front, one to the right, and one to the left of the user's viewpoint. The crates were positioned within arm's reach (1 meter away).

\subsection{Collected Metrics}
\label{sec:metrics}
We evaluated our interfaces using objective metrics (task completion times) and subjective user feedback.

\subsubsection{Objective Measures}
For objective metrics, we measured \textit{Search Time} as the time from when the coin appeared until the participant confirmed the transition to the correct environment through the interface, \textit{Retrieve Time} as the time from that transition until the coin was collected, and \textit{Deposit Time} as the time from acquisition until the coin was deposited at the starting location. Retrieve Time includes time to orient (turn toward the coin) plus time to grab (extending the arm and pinching) post-transition, so, assuming a consistent grab time, retrieve time will be lower with efficient pre-orientation. Therefore, in this experiment, retrieve time serves as an indirect proxy for a preview pattern's \textit{pre-orientation efficiency}. Another proxy is Deposit Time, which we use as a measure of the interaction techniques' \textit{return-to-home efficiency}.
For our task, we define a trial as the entire process of locating, transitioning, collecting, and returning back to deposit the coin. 

\subsubsection{Subjective Measures}
\label{sec:subjmeasures}
We collected subjective data on participants' perceptions of the interfaces in terms of workload, usability, and hedonic user experience.
We administered the questionnaires after participants had completed all trials to ensure participants would have a common baseline for comparison across all interfaces.
First, participants were asked to rank their preferences for the world-switching interfaces.
Then, participants completed the following questionnaires for each of the interfaces: the NASA Task Load Index (NASA-TLX)~\cite{hart2006tlx} assessed perceived workload across dimensions such as mental, physical, temporal, effort, performance, and frustration; the System Usability Scale (SUS)~\cite{brooke1996sus} evaluated subjective usability; the Hedonic sub-scale of the User Experience Questionnaire (UEQ) short version~\cite{schrepp2017ueqs} captured perception of design originality and aesthetics;
and the Continuity questionnaire evaluated the sense of an unbroken, coherent experience, as established in previous XR transition studies~\cite{husung2019portals}.
Before completing the task-related questionnaires, we verbally summarized the experiment to participants to ensure a consistent understanding of the experiment.
The order of interface evaluation was the same as the order in which they experienced the techniques.
At all times during subjective evaluation, a second monitor displayed videos for each interface, demonstrated by an author with clear steps, hand positions, and environmental context to aid recall.
We also included a question asking participants whether they remembered the interface, for which all participants answered in the affirmative for every interface.
The experimenters also answered any questions participants had about the techniques during this time. 

After completing the questionnaires, participants took part in an informal verbal interview where we asked: \textbf{(1)} \textit{What aspects of the transition techniques used in the experiment do you like and dislike?} \textbf{(2)} \textit{What features should be included in an interface that enables you to switch (transition) between different VR worlds?} \textbf{(3)} \textit{Did you plan your body movements (e.g., orienting yourself towards the coin to be picked up, or orienting yourself towards the chest) before transitioning to an environment? If so, which techniques supported this planning most effectively, and why?} \textbf{(4)} \textit{Do you think the features of a world-switching interface should differ based on the context and type of the transition? For example, what features would you want for previewing and transitioning to dynamic or cluttered environments?}

\subsection{Worlds Simulation Implementation and Apparatus}
\label{sec:apparatus}
Participants used the Meta Quest Pro HWD connected to a laptop via a Quest Link cable to complete the experiment.
The Meta Quest Pro HWD has a resolution of $1800\times1920$ per eye and a \ang{96} diagonal field of view.
We used Meta Quest Pro light blockers to minimize HWD light leakage.
We implemented our VR experiment using Unity 6\footnote{\url{https://unity.com/releases/unity-6}} with worlds as separate objects representing VEs in a single scene's hierarchy. Assigning different render layers to those objects managed the worlds shown and previewed. Portal previews were rendered in stereo using the Universal Rendering Pipeline's Render Objects, ensuring correct occlusion by foreground objects; WIM previews were miniature copies of VEs with spherical cutout shaders. This approach allowed simulating worlds and switching between them in our experiment,  while remaining in the same active application.
The application ran on an MSI Raider GE76 laptop equipped with an NVIDIA RTX 3080 Ti GPU (16 GB GDDR6), an Intel Core i9 CPU, and 32 GB DDR5 memory.
We ensured our experiment consistently ran at 72 frames per second.
Participants remained standing stationary in a cleared space, using only leaning and turning to perform the study task. 
The experimenter sat nearby with a laptop showing the participant's view to monitor.


\subsection{Participants}
We conducted an a priori power analysis in G*Power (v3.1.9.7) for a repeated-measures ANOVA (within-subjects) with the following parameters: $1$ group, $9$ measurements, a medium effect size ($f = .25$), $\alpha = .05$, and power $= .95$.
Based on this result and guidelines for similar experimental user studies~\cite{bergstrom2021evaluate}, we recruited
$22$ participants from our university ($7$ females, $15$ males) of ages ranging from 19 to 41 ($M = 24.7$, $SD = 4.7$),
with self-reported VR experience levels from 1 (never) to 5 (daily) 
($M = 2.7$, $SD = 1.0$).
All participants were right-handed.
All participants spoke English, walked independently, were not color blind, and had no neuropathic or physical disabilities. Participants had normal or corrected-to-normal vision. 
The experimental procedure and participant recruitment were approved by the institutional review board (IRB) of our university.

\subsection{Experiment Procedure}
\label{procedure}
Upon arrival, participants received a consent form with all study details. After we answered any of their questions, they provided consent. Next, we explained in detail the in-VR task, further explaining any ambiguities. Then, we explained the study apparatus and assisted participants in wearing it, and adjusting the lenses to fit their IPD. Once we confirmed the headset fit comfortably and provided a clear image, we guided the participant to the experiment's starting point. 

The VR experiment began in the ``Home'' environment (see \figref{fig:environments}-(f)), where a UI displayed the current interface’s name, activation steps, destination environment selection method, and confirmation/cancellation procedures, along with a looping video demonstrating proper execution. Participants were instructed to practice the interface by transitioning between any of the study environments until they felt comfortable with the interface. A dedicated blue coin was located behind the initial orientation of the participant in the home environment. Participants picked up this coin and deposited it into the chest to indicate their readiness to begin the actual study task trials (\secref{sec:task}). Participants completed 15 trials of the task with each interface to ensure sufficient exposure for post-study questionnaires and interviews, and upon finishing this trial block, the training UI reappeared to introduce the next interface. 
Participants were given the option to take a break in-between trial blocks, but no participant chose to do so.

The order of interfaces was counterbalanced across participants using an incomplete Latin square. After completing 15 trials per 9 interfaces (135 trials total),
the in-VR experiment concluded. Next, participants filled out web-based questionnaires evaluating the interfaces before participating in a verbal informal interview with the experimenter. Finally, participants filled out a demographics questionnaire. The experiment took approximately 75 minutes to complete and no participant exhibited signs of fatigue.
Participants were compensated with \$15 gift cards.

\begin{figure}[t!]
    \centering
    \includegraphics[width=1\linewidth]{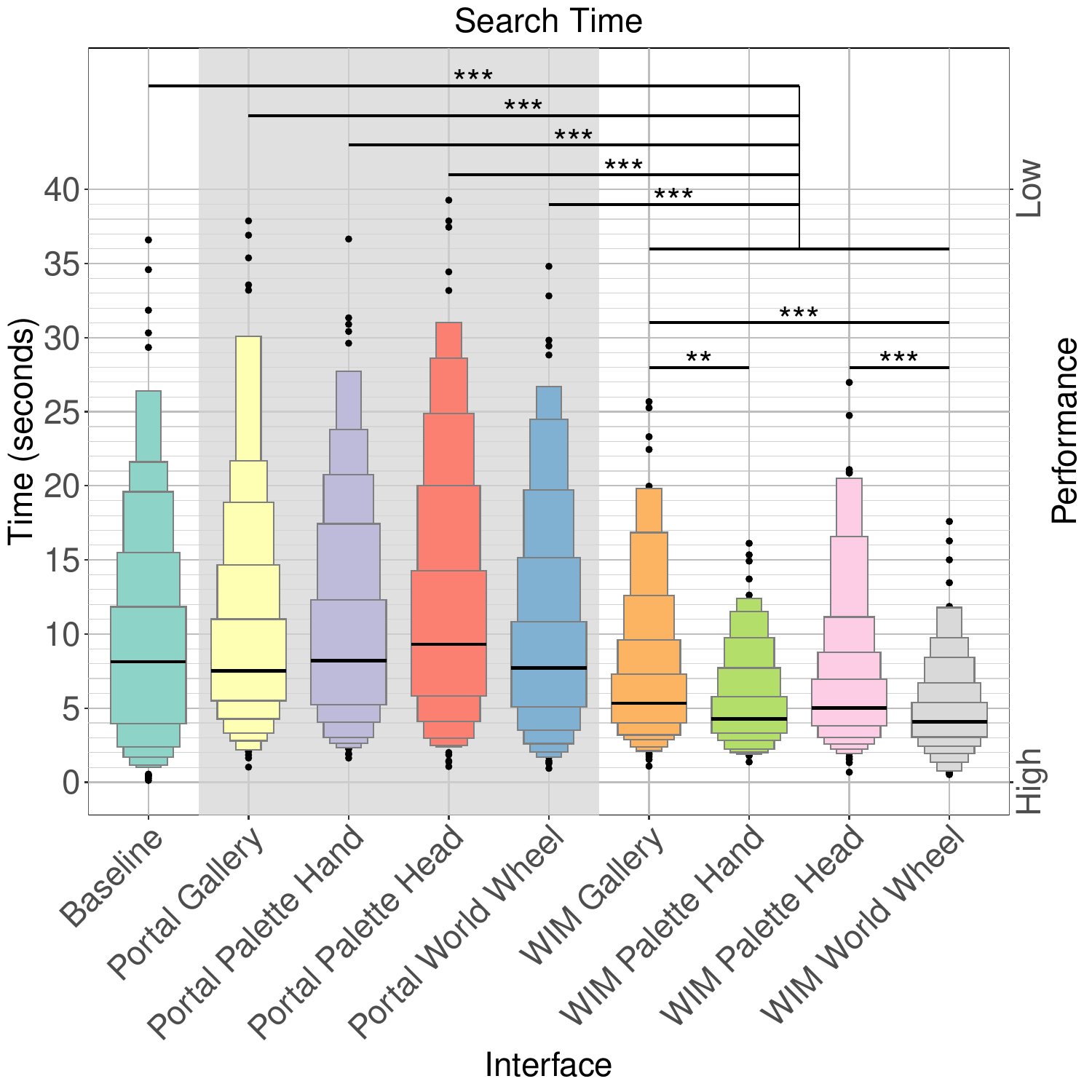}
    \caption{Search time performance across all VR world-switching interfaces. Letter-value (boxen) plots summarize distributions for search times. Lower values indicate better performance. Asterisks denote significant pairwise differences (*\,\textbf{p}<0.05, **\,\textbf{p}<0.01, ***\,\textbf{p}<0.001).}
    \label{fig:search}
    \Description{Letter-value plot showing search time in seconds across nine interface conditions: Baseline, Portal Gallery, Portal Palette Hand, Portal Palette Head, Portal World Wheel, WIM Gallery, WIM Palette Hand, WIM Palette Head, and WIM World Wheel. Each distribution is represented by progressively nested box shapes that display the median, quartiles, and additional letter values, with outliers shown as individual points. Overall performance improves from left to right, with the WIM interfaces, particularly WIM World Wheel, yielding lower search times. Horizontal significance bars with asterisks above the plots indicate multiple statistically significant differences between conditions, where more asterisks denote higher levels of significance.}
\end{figure}

\section{Results}
\label{sec:results}
In this section, we report the statistically significant results from our quantitative data.
Tables including all statistical test results can be found in the Supplementary Materials.
We performed our data analysis using R version 4.4.0 and the following packages:
\verb|dplyr|, 
\verb|ARTool|,
\verb|rstatix|,
\verb|effectsize|,
\verb|emmeans|,
\verb|ggplot2|.

\subsection{Objective Data}

We analyzed participants’ task performance across three dependent measures: Search Time, Retrieve Time, and Deposit Time.  
Each measure was analyzed with respect to its corresponding experimental factor, as described in~\secref{sec:task}: Interface (Search Time), Preview Pattern (Retrieve Time), and Interaction Technique (Deposit Time).
Extreme outliers more than four standard deviations from the mean were removed separately for each measure prior to analysis.
We assessed normality of the task time distributions using Shapiro--Wilk tests.  
All three measures deviated significantly from normality (Search Time: $W = 0.82$, $p < .001$; Retrieve Time: $W = 0.84$, $p < .001$; Deposit Time: $W = 0.68$, $p < .001$).  
Visual inspection of Q--Q plots and histograms confirmed skew and long tails, motivating the use of the nonparametric aligned rank transform procedure for subsequent analyses.  

We then averaged the data per Participant~$\times$~Condition and analyzed the resulting dataset using the Aligned Rank Transform (ART) procedure~\cite{wobbrock2011art}, implemented in the \texttt{ARTool} package in R.  
The ART method allows for standard ANOVA procedures to be applied to nonparametric data, providing robustness to violations of normality and heteroscedasticity.  
For each measure, we modeled participants’ averaged responses as a function of the primary experimental factor, with ParticipantID included as a within-subject error term.  
Overall significance of the main factor was assessed with Type~II ANOVA tests on the aligned ranks.  
Where significant main effects were found, we conducted post hoc pairwise comparisons of estimated marginal means with Bonferroni correction for multiple comparisons.  
Partial $\eta^2$ are reported as a measure of effect size.
All estimates are reported on the raw time scale as group means accompanied by 95\% confidence intervals in brackets.  
Significant pairwise differences are also visualized in the results figures.  
The Supplementary Materials provide complete ANOVA tables and pairwise test results (including degrees of freedom, t-values, etc.).


\subsubsection{Search Time}
We examined the effect of Interface on participants' Search Time.  
These data are visualized in~\figref{fig:search}.
$34$ outlier trials were excluded ($\approx 1.1\%$ of the data).  
The was a significant main effect of Interface on Search Time, $F(8, 168) = 47.03$, $\mathbf{p < 0.001}$.  
The effect size was large, with partial $\eta^2 = 0.69$ [$0.63$, $1.00$].  

Post hoc pairwise comparisons showed that all \wim{} interfaces produced significantly faster search times than \baseline{} ($8.85$~s [$7.68$, $10.0$], $p < .001$ for all):  
\wimGallery{} ($6.34$~s [$5.60$, $7.08$]),  
\wimHandPalette{} ($5.01$~s [$4.35$, $5.67$]),  
\wimHeadPalette{} ($6.00$~s [$5.20$, $6.80$]), and  
\wimSteeringWheel{} ($4.55$~s [$4.08$, $5.02$]).  
In addition, each of the four \portal{} interfaces --- \portalGallery{} ($9.20$~s [$8.09$, $10.3$]),  
\portalHandPalette{} ($9.77$~s [$8.25$, $11.3$]),  
\portalHeadPalette{} ($11.2$~s [$9.26$, $13.1$]), and  
\portalSteeringWheel{} ($9.05$~s [$7.71$, $10.4$]) --- was significantly slower than every \wim{} interface ($p < .001$ for all).  

Within the \wim{} set, \wimHandPalette{} ($5.01$~s [$4.35$, $5.67$]) was significantly faster than \wimGallery{} ($p = .010$),  
and \wimSteeringWheel{} ($4.55$~s [$4.08$, $5.02$]) was significantly faster than both \wimGallery{} ($6.34$~s [$5.60$, $7.08$]) ($p < .001$) and \wimHeadPalette{} ($6.00$~s [$5.20$, $6.80$], $p = .007$).  
No other pairwise differences reached significance.  

\begin{figure}[h!]
    \centering
    \includegraphics[width=\linewidth]{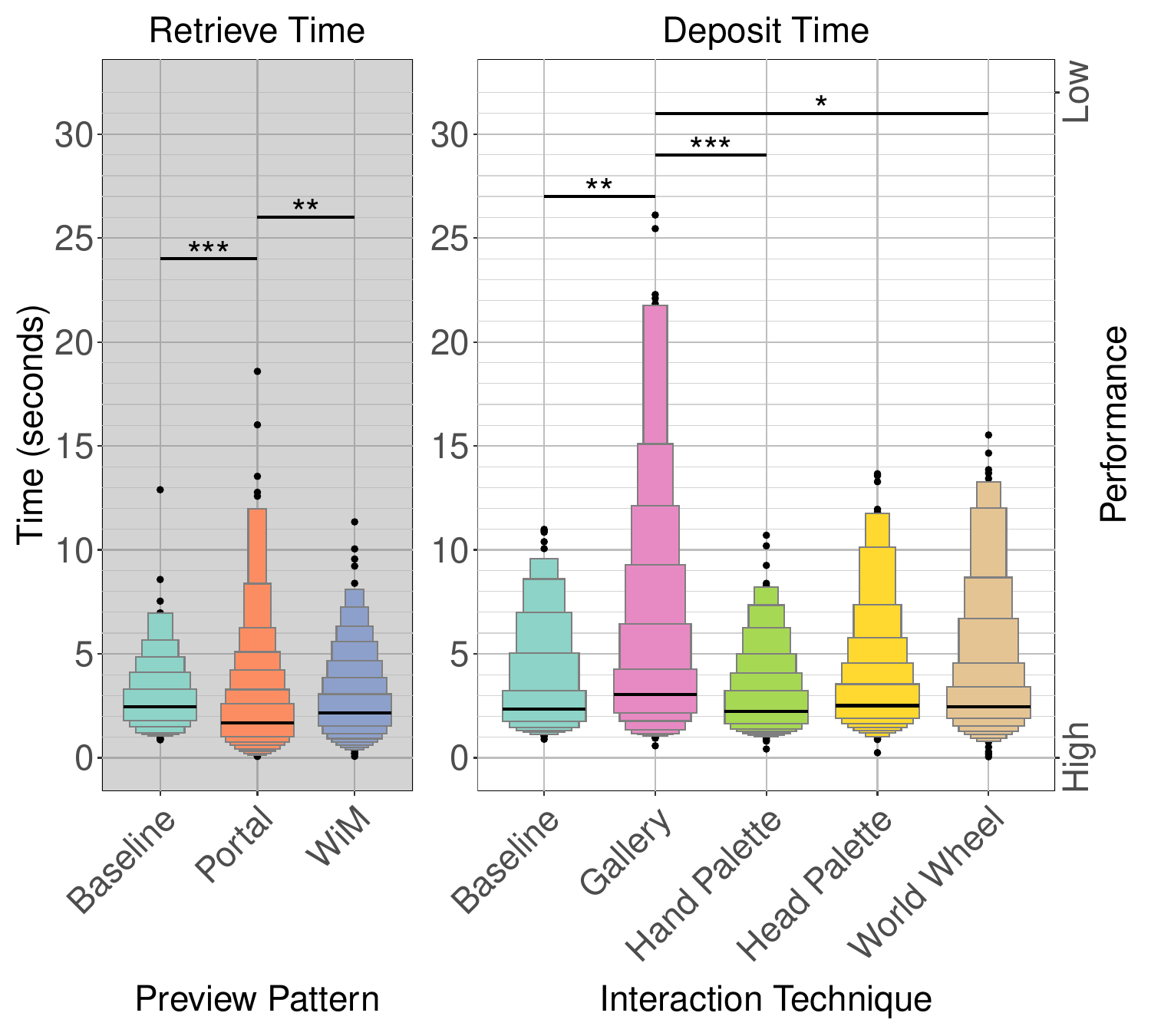}
    \caption{
    Letter-value (boxen) plots of retrieve times by preview pattern (left) and deposit times by interaction technique (right). Lower values indicate better performance. Asterisks denote significant pairwise comparisons (* $\mathbf{p < 0.05}$, ** $\mathbf{p < 0.01}$, *** $\mathbf{p < 0.001}$).
    }
    \label{fig:retrievedeposit}
    \Description{Letter-value plots of task times for two measures: Retrieve Time by Preview Pattern and Deposit Time by Interaction Technique. The left panel compares Retrieve Time across Baseline, Portal, and WIM preview patterns. Distributions show that Portal and WIM produce faster retrieval than Baseline, with significance bars above the plots marking differences. The right panel compares Deposit Time across five interaction techniques: Baseline, Gallery, Hand Palette, Head Palette, and World Wheel. The Gallery condition shows notably higher deposit times than the others, with significance bars indicating statistical differences. Outliers are shown as individual points in both panels, and performance is annotated on the vertical axis from low to high.}
    \vspace{-2ex}
\end{figure}

\subsubsection{Retrieve Time}
We examined the effect of Preview Pattern on participants' Retrieve Time.
These data are visualized in~\figref{fig:retrievedeposit}.
Outlier removal excluded $25$ trials ($\approx 0.8\%$ of the data).  
There was a significant main effect of Preview Pattern on Retrieve Time, $F(2, 42) = 14.29$, $\mathbf{p < 0.001}$.  
The effect size was large, with partial $\eta^2 = 0.40$ [$0.20$, $1.00$].  

Post hoc pairwise comparisons indicated that retrieval was significantly faster with \portal{} ($2.02$~s [$1.69$, $2.36$]) compared to both \baseline{} ($2.72$~s [$2.33$, $3.11$], $\mathbf{p < 0.001}$) and \wim{} ($2.45$~s [$2.14$, $2.75$], $\mathbf{p = 0.006}$).  
The difference between \baseline{} and \wim{} was not significant.

\subsubsection{Deposit Time}
We examined the effect of Interaction Technique on participants' Deposit Time.  
These data are visualized in~\figref{fig:retrievedeposit}.  
Outlier removal excluded $26$ trials ($\approx 0.9\%$ of the data).  
There was a significant main effect of Interaction Technique on Deposit Time, $F(4, 84) = 7.45$, $\mathbf{p < 0.001}$.  
The effect size was medium-to-large, with partial $\eta^2 = 0.26$ [$0.11$, $1.00$].  

Post hoc pairwise comparisons indicated that Deposit Time was significantly slower with \gallery{} ($3.96$~s [$3.26$, $4.66$]) compared to \baseline{} ($2.98$~s [$2.58$, $3.38$], $\mathbf{p = 0.006}$).  
Additionally, \gallery{} was significantly slower than both \handPalette{} ($2.65$~s [$2.28$, $3.02$], $\mathbf{p < 0.001}$) and \steeringWheel{} ($3.08$~s [$2.61$, $3.56$], $\mathbf{p = 0.013}$).  
No other pairwise differences reached significance.

\subsection{Subjective Data}

\begin{figure}[t!]
    \centering
    \includegraphics[width=\linewidth]{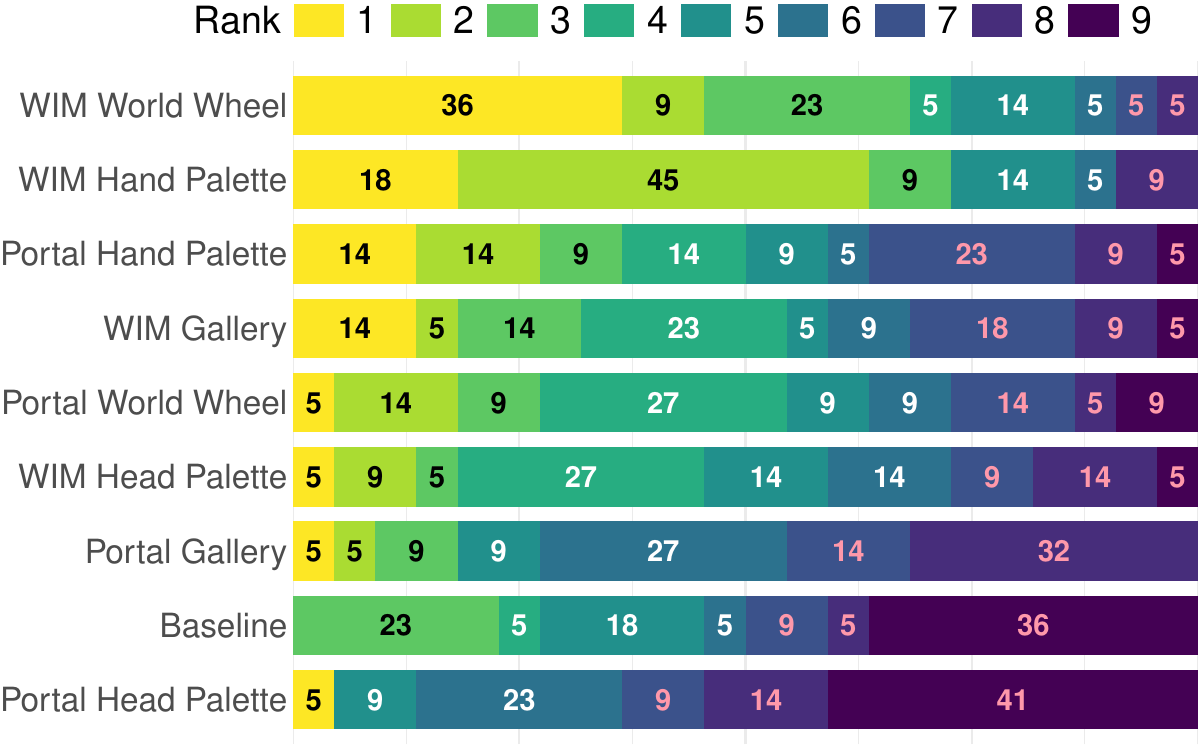}
    \caption{Ranks distribution per interface. Interfaces are ordered best-to-worst based on average ranking. Numbers within bars show percentage of participants who gave that rank (1 = highest preference, 9 = lowest). Text color indicates ranking: top 3 in black, middle 3 in white, bottom 3 in pink.}
    \label{fig:ranking}
    \Description{Stacked bar chart showing participant rankings of nine world switching interfaces from 1 (best) to 9 (worst). Each row corresponds to an interface: WIM World Wheel, WIM Hand Palette, Portal Hand Palette, WIM Gallery, Portal World Wheel, WIM Head Palette, Portal Gallery, Baseline, and Portal Head Palette,. Colored segments represent the percentage of participants assigning each rank, with yellow indicating rank 1 and dark purple rank 9. WIM World Wheel and WIM Hand Palette received the highest proportion of top rankings, while Baseline and Portal Head Palette received the highest proportion of bottom rankings. Intermediate interfaces show more evenly distributed rankings.}
    \vspace{-2ex}
\end{figure}

\begin{figure*}[t]
    \centering
    \includegraphics[width=\textwidth]{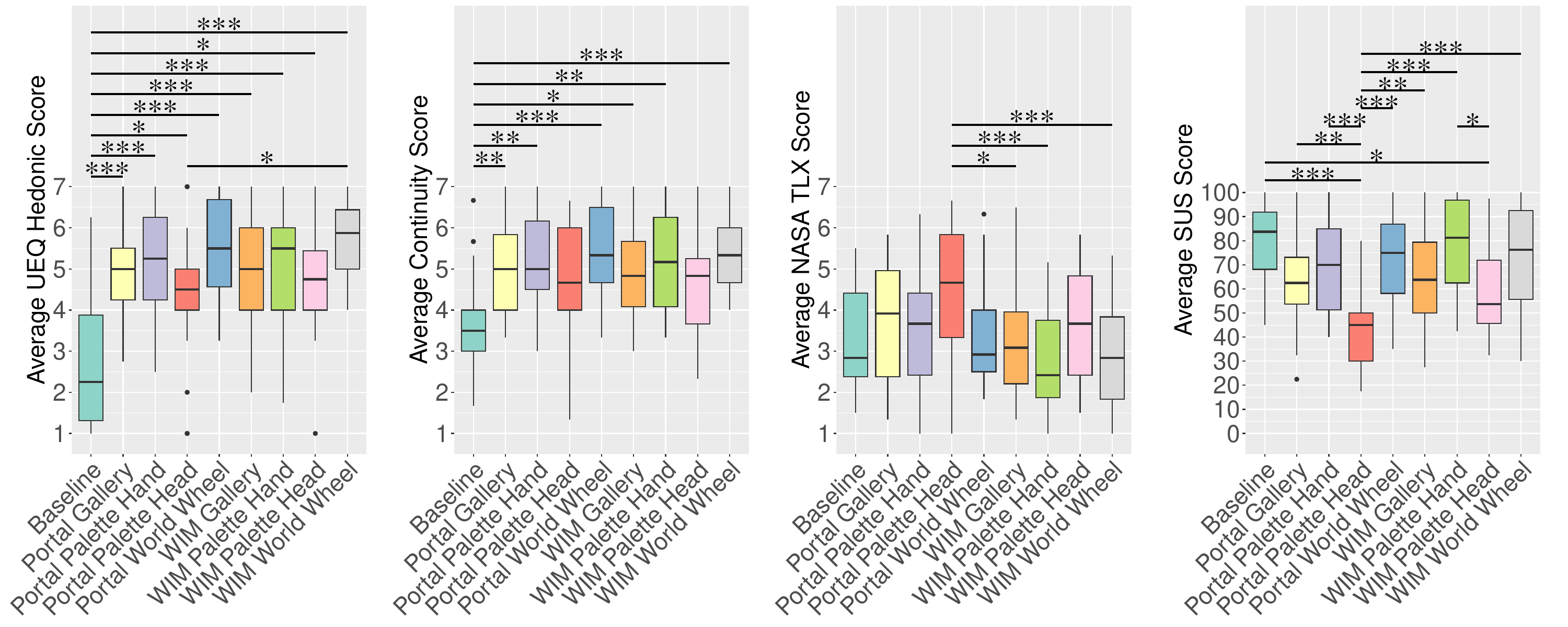}
    \Description{Four boxplots comparing subjective ratings across nine interface conditions: Baseline, Portal Gallery, Portal Palette Hand, Portal Palette Head, Portal World Wheel, WIM Gallery, WIM Palette Hand, WIM Palette Head, and WIM World Wheel. The left panel shows average UEQ Hedonic Score, where higher values indicate greater enjoyment. Several portal-based techniques, especially Portal Palette Hand and Portal World Wheel, scored significantly higher than Baseline, as indicated by multiple significance bars. The second panel from the left shows average Continuity Score, where higher values indicate smoother task flow. Again, portal-based techniques scored higher than Baseline, with significance bars showing multiple statistically significant differences. The third panel shows the average NASA TLX Score, where lower values indicate lower perceived workload. Portal-based techniques generally show lower workload than Baseline, with significance bars marking several significant differences. The right-most panel shows average System Usability Scale (SUS) score, where higher values indicate greater usability. Several WIM-based techniques scored significantly higher than Baseline, as shown by multiple significance bars above the plots. Outliers are plotted as individual points.}
    \caption{
    Subjective measures across VR world switching interfaces. From left to right, box plots summarize hedonic experience, perceived continuity, workload, and usability. Asterisks denote significant pairwise comparisons (*\,\textbf{p}<0.05, **\,\textbf{p}<0.01, ***\,\textbf{p}<0.001). Higher values indicate more favorable responses, except NASA-TLX where lower is better.
    }
    \label{fig:subjective_combined}
\end{figure*}

A chi-square test of ranking data showed a significant association between interface and ranking level, $\chi^2(64) = 154.64$, $\mathbf{p < 0.001}$, indicating non-random preference patterns.
Ranking data are visualized in~\figref{fig:ranking}.
We followed a similar analysis procedure for the rest of the subjective questionnaire data as we did for the objective data, using ART~\cite{wobbrock2011art} that modeled participants' averaged responses as a function of the Interface with ParticipantID included as a within-subject error term.
Post-hoc comparisons used Bonferroni correction for multiple comparisons.
No trials were excluded as the questionnaire responses were complete for all participants.
Questionnaire data are visualized in~\figref{fig:subjective_combined}.

\subsubsection{User Experience Questionnaire}

There was a significant main effect of Interface on participants’ hedonic quality ratings from the UEQ questionnaire, $F(8, 168) = 9.23$, $\mathbf{p < 0.001}$.
The effect size was large, with partial $\eta^2 = 0.31$ [$0.19$, $1.00$].  

Post hoc pairwise comparisons indicated that all interfaces yielded significantly higher hedonic ratings than \baseline{} ($2.61$ [$1.94$, $3.29$]): \portalGallery{} ($5.00$ [$4.46$, $5.54$], $\mathbf{p < .001}$), \portalHandPalette{} ($5.27$ [$4.71$, $5.84$], $\mathbf{p < .001}$), \portalHeadPalette{} ($4.35$ [$3.63$, $5.06$], $\mathbf{p = .014}$), \portalSteeringWheel{} ($5.49$ [$4.95$, $6.03$], $\mathbf{p < .001}$), \wimGallery{} ($4.94$ [$4.31$, $5.57$], $\mathbf{p < .001}$), \wimHandPalette{} ($5.02$ [$4.31$, $5.74$], $\mathbf{p < .001}$), \wimHeadPalette{} ($4.76$ [$4.18$, $5.35$], $\mathbf{p < .001}$), and \wimSteeringWheel{} ($5.70$ [$5.28$, $6.13$], $\mathbf{p < .001}$).  
Additionally, \wimSteeringWheel{} ($5.70$ [$5.28$, $6.13$]) received higher hedonic ratings than \portalHeadPalette{} ($4.35$ [$3.63$, $5.06$], $\mathbf{p = .011}$).  
No other pairwise contrasts were significant.

\subsubsection{Continuity}
There was a significant main effect of Interface on participants’ continuity ratings, $F(8, 168) = 5.06$, $\mathbf{p < 0.001}$.  
The effect size was medium, with partial $\eta^2 = 0.19$ [$0.08$, $1.00$].  

Post hoc pairwise comparisons indicated that continuity ratings were significantly higher for most interfaces relative to \baseline{} ($3.70$ [$3.15$, $4.24$]):  
\portalGallery{} ($5.02$ [$4.48$, $5.55$], $\mathbf{p = .007}$),  
\portalHandPalette{} ($5.17$ [$4.63$, $5.72$], $\mathbf{p < .001}$),  
\portalSteeringWheel{} ($5.38$ [$4.88$, $5.88$], $\mathbf{p < .001}$),  
\wimGallery{} ($4.82$ [$4.35$, $5.28$], $\mathbf{p = .040}$),  
\wimHandPalette{} ($5.18$ [$4.65$, $5.71$], $\mathbf{p < .001}$), and  
\wimSteeringWheel{} ($5.39$ [$4.95$, $5.84$], $\mathbf{p < .001}$).  
Differences between \baseline{} and \portalHeadPalette{} ($4.59$ [$3.92$, $5.25$]) or \wimHeadPalette{} ($4.61$ [$4.10$, $5.12$]) did not reach significance.  
No other pairwise contrasts were significant.  


\subsubsection{NASA TLX}
There was a significant main effect of Interface on participants’ subjective workload ratings (NASA TLX), $F(8, 168) = 4.22$, $\mathbf{p < 0.001}$.  
The effect size was medium, with partial $\eta^2 = 0.17$ [$0.06$, $1.00$].  

Post hoc pairwise comparisons showed \portalHeadPalette{} ($4.49$ [$3.78$, $5.21$]) rated as imposing significantly greater workload than \wimGallery{} ($3.27$ [$2.65$, $3.89$], $\mathbf{p = .028}$), \wimHandPalette{} ($2.70$ [$2.14$, $3.27$], $\mathbf{p < .001}$), and \wimSteeringWheel{} ($2.84$ [$2.26$, $3.42$], $\mathbf{p < .001}$).  
No other pairwise contrasts were significant.  

\subsubsection{System Usability Scale (SUS)}
There was a significant main effect of Interface on participants’ usability ratings from the SUS questionnaire, $F(8, 168) = 8.67$, $\mathbf{p < 0.001}$.  
The effect size was large, with partial $\eta^2 = 0.29$ [$0.18$, $1.00$].  

Post hoc pairwise comparisons indicated that \portalHeadPalette{} ($42.7$ [$35.7$, $49.8$]) received significantly lower SUS ratings than all interfaces:  
\baseline{} ($78.8$ [$71.0$, $86.5$], $\mathbf{p < .001}$),  
\portalGallery{} ($64.2$ [$55.0$, $73.4$], $\mathbf{p = .001}$),  
\portalHandPalette{} ($69.5$ [$61.2$, $77.7$], $\mathbf{p < .001}$),  
\portalSteeringWheel{} ($71.5$ [$62.9$, $80.1$], $\mathbf{p < .001}$),  
\wimGallery{} ($64.8$ [$55.2$, $74.3$], $\mathbf{p = .002}$),  
\wimHandPalette{} ($78.1$ [$69.5$, $86.7$], $\mathbf{p < .001}$),  
\wimHeadPalette{} ($59.4$ [$51.1$, $67.8$], $\mathbf{p = .011}$), and  
\wimSteeringWheel{} ($74.2$ [$64.1$, $84.3$], $\mathbf{p < .001}$).  
In addition, \wimHeadPalette{} was rated significantly lower than \wimHandPalette{} ($\mathbf{p = .010}$) and \baseline{} ($\mathbf{p = .011}$).
No other pairwise contrasts were significant.

\section{Qualitative Findings}

We used inductive thematic analysis~\cite{braun_using_2006} to analyze interview transcripts.
We coded the transcripts of participant interviews.
Then, we grouped the codes into categories and organized those categories into overarching themes.
Three authors completed each step of this process independently and met to resolve any disagreements.
In this section, we describe the main themes that emerged and summarize the codes comprising each theme.
Exemplar participant quotations for each code are provided in the Supplementary Materials.

\subsection{Theme 1: Previews Affect Orientation and Search}
This theme explains how preview modalities shaped participants’ ability to locate targets and enter a new world in a pre-oriented way. It comprises three concepts: \textit{WiM Fast Overview} ($n = 9$), \textit{Portal Pre-orientation} ($n= 6$), and \textit{WiM Pre-orientation} ($n = 12$).
Regarding \textit{WiM Fast Overview}, participants valued WiM for enabling rapid, whole-scene scanning and quick target identification. They described appreciating the ability to see everything at once and decide efficiently where to switch next.
For \textit{Portal Pre-orientation}, participants reported that the portal view supported entering the destination already facing the desired direction, reducing corrective turns after arrival.
Beyond fast search, participants valued \textit{WiM Pre-orientation}, or using the WiM to pre-plan their body orientation and reaching direction before committing to a switch. They leveraged stable landmarks (e.g., boxes, avatar position) to decide how they would be aligned upon entry.

\subsection{Theme 2: Control Fluency (Speed and Precision)}
This theme explains how input modality and control granularity shaped participants’ ability to switch quickly and accurately. It is comprised of five concepts: \textit{World Wheel Fast/Fluid} ($n = 12$), \textit{Linear Menus Imprecision} ($n = 6$), \textit{Hand-based Preference} ($n = 5$), \textit{Desire for Quick Selections} ($n = 3$), and \textit{Hand-tracking Issues ($n = 3$}).
For \textit{World Wheel Fast/Fluid}, participants described the two-handed ``wheel'' interaction as rapid and smooth to operate. They emphasized responsiveness and continuous control during selection.
Regarding \textit{Linear Menus Imprecision}, linear/scrolling menus were perceived as hard to control and overly sensitive. Participants reported difficulty stabilizing the cursor and committing to the intended world without overshoot.
Participants exhibited a \textit{Hand-based Preference} for selection over head- or ray-driven input when precise targeting and quick commitments were needed. They noted advantages for speed and confidence, especially with the dominant hand.
Participants also expressed \textit{Desire for Quick Selections}: they wanted fewer steps to commit, including a simple, reliable shortcut for common operations (e.g., returning home). They also proposed leveraging gaze to reduce input steps and speed up selection.
\textit{Hand-tracking Issues} also arose: participants reported intermittent loss of hand tracking and unintended activations, which undermined trust and slowed switching.

\subsection{Theme 3: Ergonomics and Physical Load}
This theme describes how bodily effort and posture constraints affected participants’ willingness to use particular switching techniques. It is comprised of two concepts: \textit{Head-based Cumbersome} ($n = 6$) and \textit{Physical Fatigue} ($n = 6$).
Regarding \textit{Head-based Cumbersome}, participants reported that head-driven interactions required uncomfortable neck movements and awkward postures, making them less desirable for frequent switching. They preferred techniques that could be executed with small, localized motions.
Participants described the potential for cumulative arm and shoulder \textit{Physical Fatigue} during prolonged use, especially with sustained two-handed interactions. 

\subsection{Theme 4: User and Interface Adaptation}
This theme explains how the suitability of a switching technique depended on environmental demands and user goals, and how consistent mappings and practice produced more automatic, efficient switching behaviors.
It is comprised of three concepts: \textit{Context/Task Dependence} ($n = 5$), \textit{Muscle Memory} ($n = 2$), \textit{Learning Curve} ($n=2$), and \textit{Retrieval through Portal} ($n=2$).
For \textit{Context/Task Dependence}, participants projected needing different interfaces based on the scene and task. For cluttered or dynamic environments, they favored portals to isolate content; for broad search, they preferred WiM to view all options at once; for slower, deliberate exploration, simpler baselines were acceptable.
Participants also indicated that in some tasks the world switching interface should support the ability to interact through the preview rather than fully transitioning to retrieve objects.
Participants pointed out the benefits of \textit{Muscle Memory}: with repetition, participants developed embodied shortcuts that reduced cognitive effort.
Participants also experienced a \textit{Learning Curve}, or an initial period of skill acquisition to achieve reliable input and smooth control.
Last, participants indicated that in some tasks the world switching interface should support \textit{Retrieval through Portal}, or the ability to interact through the preview rather than fully transitioning to retrieve objects.

\section{Discussion}
\label{sec:discussion}
This section integrates our quantitative and qualitative findings into actionable guidance for VR world switching.
We present design recommendations for researchers and practitioners interested in building world switching interfaces as well as directions for future related research.

\subsection{Task-Dependent Advantages for Preview Patterns}
In scenarios where the user's task is distributed across worlds, previews of worlds can improve users' task performance.
Even in interfaces that do not provide significant advantages in performance, previews are still largely preferred by participants, provide a positive hedonic experience, and increase perceptions of continuity.
WiM previews supported faster visual search by giving users a global overview~\cite{stoakley1995virtual}, while portal previews supported faster action on arrival by placing users in a pre-oriented first-person view\cite{huang2025preview, elvezio2017preorient}. Participants also reported mentally pre-orienting with WiMs~\cite{elvezio2017preorient}, but portals more consistently reduced post-transition correction. Together, these findings indicate that preview breadth (WiM) reduces the \emph{search} phase, and preview continuity (portal) reduces the \emph{retrieve} phase upon arrival.
In the interviews, participants mentioned that different scenarios would lend themselves to different preview patterns. For example, portals were favored for cluttered or dynamic environments where isolation and immediate alignment matter; WiMs were favored for rapid scanning across many candidates; simpler lists can work when time pressure is low.

\subsection{Advantages of Direct Interactions}
The return-to-home interaction was best supported by direct, low-friction hand interactions. The \gallery{} interfaces, which required fine scrolling and multiple micro-adjustments, slowed performance and increased control overhead. Hand-centric selection led to better usability.
Participants also noted the efficiency advantages of spatially-consistent interfaces for which they developed muscle memory~\cite{scarr2013spatialmemory} during the experiment.


Participants identified an opportunity to further streamline interaction: manipulation through previews. For tasks involving simple actions like object retrieval, this could eliminate the transition-retrieve-return sequence entirely, potentially offering substantial efficiency gains. This aligns with emerging work on interaction through portals~\cite{ablett2023point, lisle2022clean} and suggests that making previews directly interactive, not just viewable, represents a promising direction for reducing interaction overhead in world switching interfaces.

\subsection{Comfort and Ergonomics of Interaction}
Head-driven selection introduced neck strain and led to high workload and low usability. Two-handed controls (i.e., \handPalette{} and \steeringWheel{}) were perceived as fast and engaging but became tiring over longer runs due to having to extend both arms~\cite{palmeira2024gorilla}. 
It is important to account for the role of input modality when interpreting our results. All interfaces used hand tracking, which prior work has shown is preferred for direct touch interactions~\cite{luong2023controllershands} but can suffer from tracking instability. Controller-based implementations might yield different relative performance, particularly for techniques requiring sustained or precise input. Future work should evaluate these interfaces across modalities to distinguish modality-specific effects from general design principles.

\subsection{Design Recommendations and Future Research Directions}
\label{sec:designrecsandfutureresearch}
Based on these insights, we offer the following design and future research recommendations (R) for researchers and practitioners developing world switching interfaces:

\noindent
\textbf{R.1 Match preview pattern to the user’s task.}
\begin{itemize}
[leftmargin=1em]
    \item \textbf{Recommendations:} Provide both WiM and Portal preview modes, displayed simultaneously or alternately selectable.
    \item \textbf{Future research} should explore which preview patterns best support different kinds of tasks and environments as well as interfaces that adapt based on immediate user needs~\cite{lai2025adaptique}.
\end{itemize}

\noindent
\textbf{R.2 Use direct, body-based interaction for routine navigation.}
\begin{itemize}
[leftmargin=1em]
    \item \textbf{Recommendations:} Prefer Hand Palette-style interactions over Gallery scrolling for routine transitions, and provide spatially consistent methods for users to build muscle memory for quickly selecting to the desired world.
    \item \textbf{Future research} should explore how different world layouts and orderings can affect user performance and experience in different task contexts. For example, when frequently toggling between a limited set of worlds, the user may prefer world options to be ordered by most recent use to support quick-back shortcuts.
\end{itemize}

\noindent
\textbf{R.3 Design for low-effort posture and sustainable motion.}
\begin{itemize}
[leftmargin=1em]
    \item \textbf{Recommendations}: Favor localized hand motions and avoid head-driven selection for frequent use.
    \item \textbf{Future research} should build upon or create novel interfaces that leverage other modalities or interaction patterns (e.g., gaze + pinch~\cite{pfeuffer2017gazepinch}) to speed up world selection or enable easy interaction through the previews themselves~\cite{ablett2023point, lisle2022clean}).
\end{itemize}

\subsection{Limitations}

Our study has limitations that point to promising directions for future work. 
This work constitutes an exploratory effort to inspire further research on efficient VR world switching interfaces.
We do not claim that the VR transition interfaces and designs (see~\figref{fig:teaser}) evaluated here are the only or optimal solutions for seamless transitions. 
Methodologically speaking, a larger and more diverse sample could provide deeper insights into accessibility, adaptability, and overall effectiveness of the presented interfaces.
Additionally, as described in~\secref{sec:subjmeasures}, questionnaires were administered after participants completed all tasks with all interfaces. While this ensures a common frame of reference, administering questionnaires in VR immediately after each interface interaction may provide a more experience-proximal measure of each construct.

As mentioned in~\secref{sec:intro}, our experimental task (coin collection) was abstract compared to nuanced real-world workflows.
Specifically, our search-and-retrieve task serves the purpose of evaluating world-switching interfaces and fits with the \textit{(f) Cross-Application Interoperability} use case (see~\secref{sec:use-cases}), similar to a `copy-and-paste' task.
Considering that this is an initial exploration, future work should expand the evaluation to include additional contexts with varying workloads (i.e., more complex tasks and environments), such as the others we outlined in~\secref{sec:use-cases}.
Moreover, the worlds participants had to switch between had similar arrangements of crates, and increasing the distinctiveness between worlds and varying the search target object and placement could further emphasize the advantages and disadvantages of the different world switching techniques, or reveal additional user needs.
Future work should also evaluate world switching interfaces in non-constrained experimental setups, especially ones resembling consumer VR applications~\cite{maslych2024research}.
Moreover, we plan to explore the applicability of our interfaces in realistic workflows through modifications at the OS level to support switching between different VR applications. 

Last, participants who never used VR ($n = 2$) found it challenging to quickly get used to the interfaces due to unfamiliarity with hand tracking, suggesting the need for further investigation into learning curves and usability. Since some of our interfaces require bimanual interaction, future work should explore alternative techniques that provide interaction metaphors when both hands are occupied or to derive additional techniques that keep at least one hand free.
Other modalities (e.g., eye gaze, voice) ought to be investigated to assess performance tradeoffs and potential increased accessibility.

\vspace{-2mm}
\subsection{Future Work}

Beyond the directions for future work that build directly on our work (as mentioned in~\secref{sec:designrecsandfutureresearch}), we see additional potential for interesting research related to world switching.
First, we acknowledge that exploring interfaces that enable users to efficiently transition between multiple independent immersive applications diverges somewhat from prevailing emphases in VR research, which typically focus on maximizing user attention or presence within a single virtual environment~\cite{skarbez2017surveypresence}.
It is likely that a fast transition out of one world will lead to a break in presence.
Future research could consider methods for maintaining a user's sense of presence in a world they have left.
Another such direction relates to the display and control of world switching interfaces.
Future iterations could incorporate visual affordances to enhance usability and feedback (e.g., confirmation and cancel icons).
Our interfaces displayed the world options on a 2D plane, and there may be opportunities to leverage 3D interaction for the world selection process (e.g.,~\cite{husung2019portals, pointecker2022bridging}).
Additionally, other gestures may prove suitable for certain world switching interactions, such as ones inspired by app- and workspace- switching swipe gestures for mobile devices and desktops.
Last, our approach to scrolling was adjusted based on pilot studies, and the fine-tuning of other parameters such as layout configurations and speed thresholding
could enhance the interfaces' efficiency. These aspects warrant a future systematic exploration.

We used one transition effect (a short, simple fade-in fade-out) in our study.
In task-based environments, users prefer short transition effects~\cite{feld2024simple}, but the literature has demonstrated positive outcomes associated with other transition designs that may be longer~\cite{pointecker2022bridging, pointecker2024replica, valkov2017smooth}.
Other work suggests transition design can have an effect on cognitive performance, of special relevance to world switching for productivity purposes~\cite{GottsackerResidue}.
Future work should include different transition designs (e.g., blending or other spatial transitions~\cite{pointecker2024replica, bruderportal, pointecker2022bridging}) to observe their effects on the performance of world switching interactions.
Relatedly, our evaluation focused on VR, and future work should explore these interfaces across the MR continuum~\cite{milgramTaxonomyMixedReality1994}, which can leverage foundational work on transitional interfaces to smooth environmental change across realities, maintain spatial orientation and reduce disorientation, as well as support cognitive and narrative continuity~\cite{pavavimol2025transitions, billinghurst_magicbook_2001}.
Further, research showed performance in non-VR task-switching contexts is affected by factors such as task complexity and how related tasks are to each other~\cite{trafton2007taskinterruptions}.
It is therefore worth examining how different levels of complexity and similarity in spatial tasks might affect users' performance with different interfaces, and what additional techniques are needed to help best manage such scenarios.

Our work focuses on quickly switching between spatial tasks in VR.
However, there is a growing body of work on using VR to multitask for traditional desktop-based knowledge work~\cite{pavanatto2021monitors}.
We recommend conducting studies on world switching in productivity scenarios, e.g., using spatialized cues to enable efficient mental context recovery~\cite{bahnsen_augmented_2024} when switching between immersive workspaces~\cite{lisle2020thinkspaces}.
Other avenues for future work include world switching interfaces that go beyond just switching the user's environment.
For example, a user's \textit{embodiment} may change as they transition from one world to another --- their avatar may change as they transition from a work meeting to a social gathering, or their virtual toolset can change as they use different applications.
Finally, we recommend that future work explore psychological aspects of world switching paradigms and presence distributed across multiple worlds simultaneously~\cite{ablettSimultaneousPresenceContinuum2025}, e.g., effects of spatial multitasking on users' attention and memory.

\section{Conclusion}
In this paper, we presented a novel interaction model for efficiently switching between virtual worlds.
We developed and evaluated eight world switching interfaces combining two preview patterns (WiM and Portal) with four interaction techniques (Hand Palette, Head Palette, World Wheel, and Gallery). Our empirical evaluation showed significant impacts of interface features on performance, cognitive load, and user experience for an object collection task distributed across distinct VEs.
WiM previews were particularly effective for providing rapid spatial awareness while Portal previews enabled effective pre-orientation. 
Direct, body-based interaction techniques such as the Hand Palette and World Wheel showed strong user preference and usability, especially when combined with the WiM preview.
Based on these insights, we offer recommendations for using our techniques and designing new effective interfaces in this context. Our findings provide a foundation for developing intuitive, efficient, and enjoyable world-switching methods in VR. Future research should explore additional interaction modalities, refine existing metaphors, and evaluate these techniques within broader realistic VR contexts.


\begin{acks}
This work was supported in part by the U.S. National Science Foundation (NSF) Award \#2506427; NSF Award \#2235066, Collaborative Research: CCRI: Grand: Virtual Experience Research Accelerator (VERA), under NSF Program Director Han-Wei Shen; the Office of Naval Research under Award Numbers N00014-25-1-2159 and N00014-25-1-2245 (Dr. Peter Squire, Code 34); and the U.S. Army Research Lab Award W912CG2320004.
We also acknowledge AdventHealth for their support of Prof. Welch via their Endowed Chair in Healthcare Simulation.
\end{acks}

\bibliographystyle{ACM-Reference-Format}
\bibliography{00_references}

\onecolumn
\newpage

\appendix
\section{Supplementary Materials}

\subsection{Gallery Interaction Implementation Details}
During pilot testing we found that isomorphic linear scrolling through the options was sometimes difficult for participants, hence, to support both fine- and coarse-grained control in the same scrolling motion, we used a non-isomorphic scrolling function.
We tuned the scrolling parameters through and describe our approach below. 
Let \( p_x(t) \) denote the hand's horizontal position at time \( t \) and \(\Delta t\) be the time elapsed between successive measurements. The hand's horizontal velocity $v_x(t)$ is computed as:
\begin{equation}
\scalebox{0.95}{$
v_x(t) = \frac{p_x(t) - p_x(t-\Delta t)}{\Delta t}
$}
\end{equation}

Additionally, the vertical velocity component $v_y(t)$ is calculated the same way. Scrolling only occurs when horizontal movement dominates vertical movement ($|v_y| \leq |v_x|$).

We defined lower and upper velocity thresholds \( v_l \) and \( v_u \), with corresponding gain multipliers \( m_l \) and \( m_u \). The dynamic gain multiplier, \(\alpha(v_x)\), is defined as:
\begin{equation}
\scalebox{0.95}{$
\alpha(v_x) = \min\left\{ m_u, \, \max\left\{ m_l, \; m_l + \frac{m_u - m_l}{v_u - v_l}\left(|v_x| - v_l\right) \right\} \right\}
$}
\end{equation}

In our implementation, we used the following parameter values (in m/s): $v_l = 0$, $v_u = 10$, $m_l = 0.05$, and $m_u = 0.1$. With \( s \) being the gesture-specific scrolling speed factor ($s = 1.0$ in our implementation), the incremental cursor displacement \(\Delta x\) is given by:
\begin{equation}
\scalebox{0.95}{$
\Delta x = s\, v_x\, \alpha(v_x)
$}
\end{equation}

If the current cursor position is \( x_{\text{current}} \) and the valid range of motion is bounded by \( x_{\text{start}} \) and \( x_{\text{end}} \), the updated cursor position $x_{\text{new}}$ is computed as:

\begin{equation}
\scalebox{0.95}{$
x_{\text{new}} = \operatorname{clamp}\Bigl(x_{\text{current}} + \Delta x,\; x_{\text{start}},\; x_{\text{end}}\Bigr)
$}
\end{equation}

Additionally, we implemented a snapping mechanism for precise selection. When the hand's horizontal velocity falls below a threshold ($|v_x(t)| < 0.05$), the cursor position is linearly interpolated toward the nearest discrete snap point with a weight of $0.5$. The snap point $x_{\text{target}}$ is calculated as:
\begin{equation}
\scalebox{0.95}{$
x_{\text{target}} = \text{round}\left(\frac{x_{\text{new}} - x_{\text{start}}}{\delta_{\text{scroll}}}\right) \cdot \delta_{\text{scroll}} + x_{\text{start}}
$}
\end{equation}
where $\delta_{\text{scroll}}$ is the interval spacing between interface elements.
Releasing the pinch with an upward slide confirms the transition, while a downward release cancels it and disables the interface.
While the pinch moves vertically, the cursor remains in place.

\newpage

\subsection{Instructional User Interface}
\begin{figure}[h!]
    \centering
    \includegraphics[width=.6\columnwidth,clip,trim=0 80 0 0]{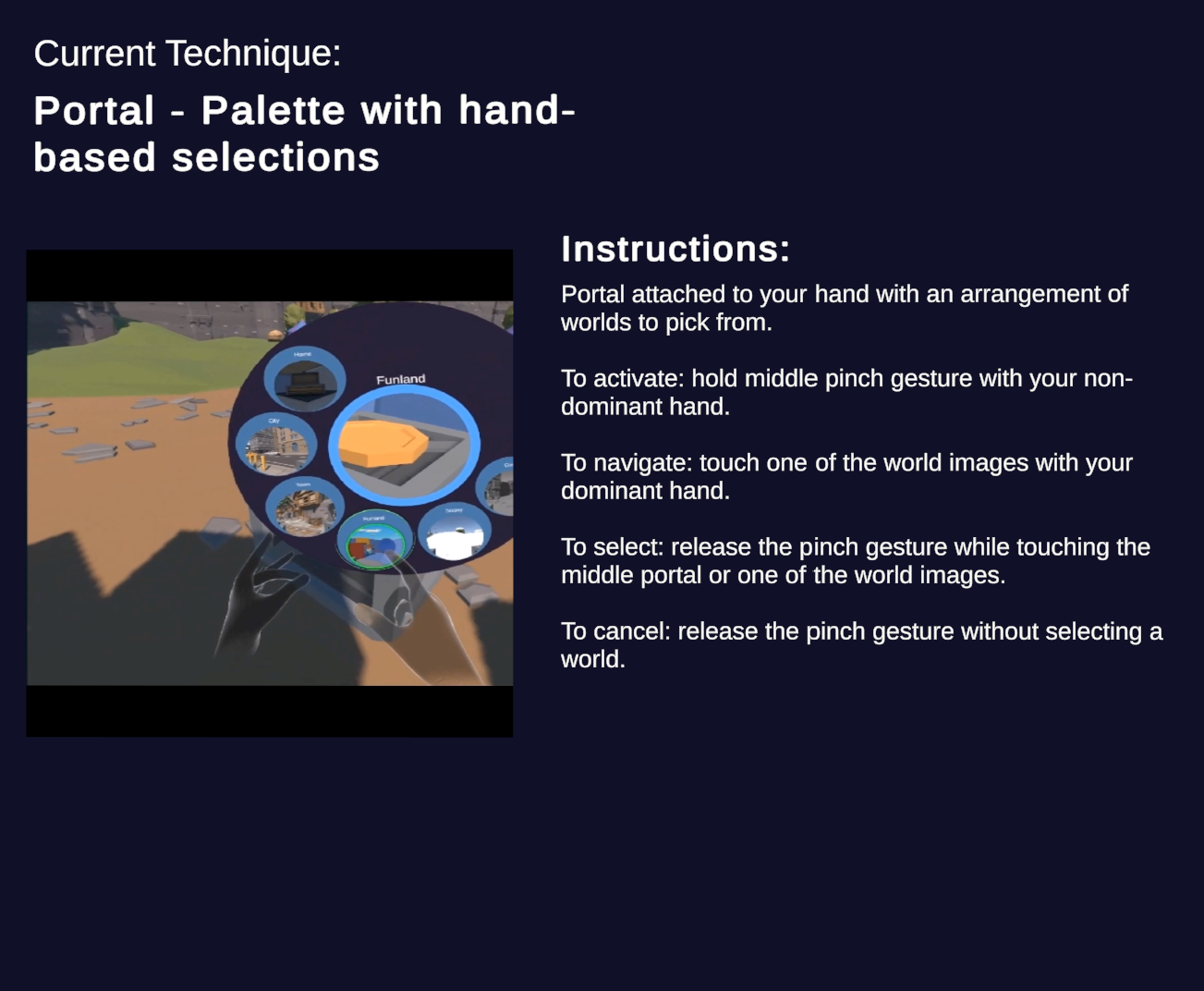}
    \caption{UI shown in the ``Home'' environment containing text instructions and a looping video demonstration of the technique. 
    }
    \label{fig:ui}
    \Description{Instructional screen for the Portal - Palette with hand-based selections technique. The screen shows a portal menu attached to the user’s hand with circular previews of available worlds. Instructions explain that to activate the menu, the user holds a middle pinch gesture with their non-dominant hand. To navigate, they touch one of the world images with their dominant hand. To select, they release the pinch gesture while touching the middle portal or one of the world images. To cancel, they release the pinch gesture without selecting a world.}
\end{figure}

\newpage

\subsection{Qualitative Analysis Details}
\vspace{-2ex}
\begin{table}[h!]
\centering
\footnotesize
\caption{Summary of qualitative analysis results with multiple exemplar quotations and PID attribution.}
\vspace{-2ex}
\begin{tabular}{p{2.6cm} p{3cm} p{0.8cm} p{6.3cm} p{0.8cm}}
\toprule
\textbf{Theme} & \textbf{Code} & \textbf{\# PIDs} & \textbf{Example quotation(s)} & \textbf{PID} \\
\midrule

\multirow[t]{3}{=}{\textbf{Previews Affect Orientation and Search}}
  & WiM fast overview & 9 
  & ``the WiM was the quickest way to get an immediate glance of the environment\ldots quick scan, look for the coin.'' 
  & 16 \\

  &  &  & ``for the WiM, it was super easy [to find] the coin.'' & 21 \\

  &  &  & ``any time that I was able to see all the boxes at once\ldots that was the one I preferred the most.'' & 19 \\
\midrule

\multirow[t]{2}{=}{}
  & Portal pre-orientation & 6 
  & ``[the] portal isolates the field of view\ldots you are already oriented as you go into the new environment.'' 
  & 21 \\

  &  &  & ``for the portal, you are already oriented as you went into that new environment.'' & 3 \\
\midrule

\multirow[t]{3}{=}{}
  & WiM pre-orientation & 12 
  & ``regardless of the orientation of the overhead view\ldots I was able to know immediately which direction to reach\ldots because my avatar’s position.'' 
  & 9 \\

  &  &  & ``with the WiM\ldots I could orient myself without visually looking.'' & 16 \\

  &  &  & ``with the world miniature\ldots I know exactly which box [was] to my right.'' & 22 \\

\midrule
\midrule

\multirow[t]{3}{=}{\textbf{Control Fluency (Speed and Precision)}}
  & Steering wheel fast/fluid & 12
  & ``steering wheel [was] really fast, really responsive, and it felt\ldots natural.''
  & 16 \\

  &  &  & ``the steering wheel was really, really fun [and] more fluid.'' & 21 \\
  &  &  & ``I liked the steering wheel best, it was the fastest for sure.'' & 7 \\
\midrule

\multirow[t]{2}{=}{}
  & Linear menus imprecision & 6
  & ``I didn’t love the \,[linear] one\ldots took a while to get the gesture\ldots hard to control.''
  & 22 \\

  &  &  & ``the linear menu\ldots too much user input in every single one.'' & 12 \\
\midrule

\multirow[t]{2}{=}{}
  & Hand-based preference & 5
  & ``moved faster through UIs\ldots with my dominant hand.''
  & 8 \\

  &  &  & ``liked the ones where I did more hand selection.'' & 21 \\
\midrule

\multirow[t]{4}{=}{}
  & Desire for quick selections & 3
  & ``shortcut to home\ldots would be nice.''
  & 20 \\

  &  &  & ``as soon as you touch the world, you would just\ldots go there.'' & 9 \\

  &  &  & ``gaze tracking\ldots quick gaze flicks\ldots may make selection faster.'' & 11 \\
\midrule

\multirow[t]{3}{=}{}
  & Hand-tracking issues & 3
  & ``it would just like bloop to so many environments because it would lose the hand tracking on my right hand.''
  & 21 \\

  &  &  & ``the flakiness of the tracking\ldots sometimes it would select when I didn’t want it to.'' & 20 \\

  &  &  & ``the sensitivity of recognition\ldots need to modulate pinches so the system would register them.'' & 16 \\

\midrule
\midrule

\multirow[t]{3}{=}{\textbf{Ergonomics \& Physical Load}}
  & Head-based cumbersome & 6
  & ``head-mounted stuff was also the hardest\ldots I had to put my head in weird positions.'' & 22 \\
  &  &  & ``head-based methods were difficult to navigate.'' & 11 \\
  &  &  & ``having to physically move my neck around was cumbersome.''
  & 8 \\
\midrule

\multirow[t]{3}{=}{}
  & Physical fatigue & 6
  & ``two-handed\ldots was really nice\ldots but tiring.''
  & 22 \\

  &  &  & ``with the steering wheel\ldots towards that last trial [my arm] was getting a little bit sore.'' & 3 \\

  &  &  & ``gorilla arms\ldots exhausting after a while.'' & 17 \\

\midrule
\midrule

\multirow[t]{3}{=}{\textbf{User and Interface Adaptation}}
  & Context/task dependence & 5
  & ``portal would probably work better for cluttered environments\ldots isolates the field of view.''
  & 21 \\

  &  &  & ``for a lot of clutter\ldots [the] world miniature\ldots wouldn’t perform so well\ldots portal will work better.'' & 9 \\

  &  &  & ``shows all at the same time'' for rapid scanning. & 22 \\
\midrule

\multirow[t]{2}{=}{}
  & Muscle memory & 2
  & ``going home\ldots was always kind of this L shape\ldots [it] started to muscle memory in.''
  & 17 \\

  &  &  & ``repeated use led to actions becoming automatic.'' & 3 \\
\midrule

\multirow[t]{2}{=}{}
  & Learning curve & 2
  & ``there was a skill to it\ldots learning curve\ldots how you pinch\ldots so that the headset recognizes it.''
  & 16 \\

  &  &  & ``you develop your own strategy'' & 20 \\
\midrule

\multirow[t]{2}{=}{}
  & Retrieval through portal & 2
  & ``it should support [the ability to] reach through the portal\ldots just grab it''
  & 17 \\

\bottomrule
\end{tabular}
\label{tab:qual}
\end{table}

\newpage
\newpage

\subsection{Statistical Results}



\begin{table}[h]
\centering
\scriptsize
\setlength{\tabcolsep}{5pt}
\caption{Pairwise post hoc comparisons of \emph{Search Time} between Interfaces (Bonferroni adjustment for 36 tests). $p$-values are bolded if $< 0.05$.}
\begin{tabular}{lrrrrr}
\toprule
Contrast & Estimate & SE & df & $t$ & $p$ \\
\midrule
\baseline{} -- \portalGallery{} & $-11.82$ & $9.10$ & $168$ & $-1.299$ & $1.000$ \\
\baseline{} -- \portalHandPalette{} & $-9.59$ & $9.10$ & $168$ & $-1.054$ & $1.000$ \\
\baseline{} -- \portalHeadPalette{} & $-23.50$ & $9.10$ & $168$ & $-2.583$ & $0.383$ \\
\baseline{} -- \portalSteeringWheel{} & $-1.91$ & $9.10$ & $168$ & $-0.210$ & $1.000$ \\
\baseline{} -- \wimGallery{} & $47.32$ & $9.10$ & $168$ & $5.201$ & $\mathbf{< 0.001}$ \\
\baseline{} -- \wimHandPalette{} & $81.09$ & $9.10$ & $168$ & $8.914$ & $\mathbf{< 0.001}$ \\
\baseline{} -- \wimHeadPalette{} & $58.09$ & $9.10$ & $168$ & $6.386$ & $\mathbf{< 0.001}$ \\
\baseline{} -- \wimSteeringWheel{} & $92.68$ & $9.10$ & $168$ & $10.188$ & $\mathbf{< 0.001}$ \\
\portalGallery{} -- \portalHandPalette{} & $2.23$ & $9.10$ & $168$ & $0.245$ & $1.000$ \\
\portalGallery{} -- \portalHeadPalette{} & $-11.68$ & $9.10$ & $168$ & $-1.284$ & $1.000$ \\
\portalGallery{} -- \portalSteeringWheel{} & $9.91$ & $9.10$ & $168$ & $1.089$ & $1.000$ \\
\portalGallery{} -- \wimGallery{} & $59.14$ & $9.10$ & $168$ & $6.500$ & $\mathbf{< 0.001}$ \\
\portalGallery{} -- \wimHandPalette{} & $92.91$ & $9.10$ & $168$ & $10.213$ & $\mathbf{< 0.001}$ \\
\portalGallery{} -- \wimHeadPalette{} & $69.91$ & $9.10$ & $168$ & $7.685$ & $\mathbf{< 0.001}$ \\
\portalGallery{} -- \wimSteeringWheel{} & $104.50$ & $9.10$ & $168$ & $11.487$ & $\mathbf{< 0.001}$ \\
\portalHandPalette{} -- \portalHeadPalette{} & $-13.91$ & $9.10$ & $168$ & $-1.529$ & $1.000$ \\
\portalHandPalette{} -- \portalSteeringWheel{} & $7.68$ & $9.10$ & $168$ & $0.844$ & $1.000$ \\
\portalHandPalette{} -- \wimGallery{} & $56.91$ & $9.10$ & $168$ & $6.256$ & $\mathbf{< 0.001}$ \\
\portalHandPalette{} -- \wimHandPalette{} & $90.68$ & $9.10$ & $168$ & $9.968$ & $\mathbf{< 0.001}$ \\
\portalHandPalette{} -- \wimHeadPalette{} & $67.68$ & $9.10$ & $168$ & $7.440$ & $\mathbf{< 0.001}$ \\
\portalHandPalette{} -- \wimSteeringWheel{} & $102.27$ & $9.10$ & $168$ & $11.242$ & $\mathbf{< 0.001}$ \\
\portalHeadPalette{} -- \portalSteeringWheel{} & $21.59$ & $9.10$ & $168$ & $2.373$ & $0.675$ \\
\portalHeadPalette{} -- \wimGallery{} & $70.82$ & $9.10$ & $168$ & $7.785$ & $\mathbf{< 0.001}$ \\
\portalHeadPalette{} -- \wimHandPalette{} & $104.59$ & $9.10$ & $168$ & $11.497$ & $\mathbf{< 0.001}$ \\
\portalHeadPalette{} -- \wimHeadPalette{} & $81.59$ & $9.10$ & $168$ & $8.969$ & $\mathbf{< 0.001}$ \\
\portalHeadPalette{} -- \wimSteeringWheel{} & $116.18$ & $9.10$ & $168$ & $12.771$ & $\mathbf{< 0.001}$ \\
\portalSteeringWheel{} -- \wimGallery{} & $49.23$ & $9.10$ & $168$ & $5.411$ & $\mathbf{< 0.001}$ \\
\portalSteeringWheel{} -- \wimHandPalette{} & $83.00$ & $9.10$ & $168$ & $9.124$ & $\mathbf{< 0.001}$ \\
\portalSteeringWheel{} -- \wimHeadPalette{} & $60.00$ & $9.10$ & $168$ & $6.595$ & $\mathbf{< 0.001}$ \\
\portalSteeringWheel{} -- \wimSteeringWheel{} & $94.59$ & $9.10$ & $168$ & $10.398$ & $\mathbf{< 0.001}$ \\
\wimGallery{} -- \wimHandPalette{} & $33.77$ & $9.10$ & $168$ & $3.712$ & $\mathbf{0.010}$ \\
\wimGallery{} -- \wimHeadPalette{} & $10.77$ & $9.10$ & $168$ & $1.184$ & $1.000$ \\
\wimGallery{} -- \wimSteeringWheel{} & $45.36$ & $9.10$ & $168$ & $4.987$ & $\mathbf{< 0.001}$ \\
\wimHandPalette{} -- \wimHeadPalette{} & $-23.00$ & $9.10$ & $168$ & $-2.528$ & $0.446$ \\
\wimHandPalette{} -- \wimSteeringWheel{} & $11.59$ & $9.10$ & $168$ & $1.274$ & $1.000$ \\
\wimHeadPalette{} -- \wimSteeringWheel{} & $34.59$ & $9.10$ & $168$ & $3.802$ & $\mathbf{0.007}$ \\
\bottomrule
\end{tabular}
\end{table}

\begin{table}[h]
\centering
\scriptsize
\setlength{\tabcolsep}{5pt}
\caption{Pairwise post hoc comparisons of \emph{Retrieve Time} between Preview Patterns (Bonferroni adjustment for 3 tests). $p$-values are bolded if $< 0.05$.}
\begin{tabular}{lrrrrr}
\toprule
Contrast & Estimate & SE & df & $t$ & $p$ \\
\midrule
None -- \portal{} & $15.09$ & $2.85$ & $42$ & $5.298$ & $\mathbf{< 0.001}$ \\
None -- \wim{} & $5.77$ & $2.85$ & $42$ & $2.027$ & $0.147$ \\
\portal{} -- \wim{} & $-9.32$ & $2.85$ & $42$ & $-3.271$ & $\mathbf{0.006}$ \\
\bottomrule
\end{tabular}
\end{table}


\begin{table}[h]
\centering
\scriptsize
\setlength{\tabcolsep}{5pt}
\caption{Pairwise post hoc comparisons of \emph{Deposit Time} between Interaction Techniques (Bonferroni adjustment for 10 tests). $p$-values are bolded if $< 0.05$.}
\begin{tabular}{lrrrrr}
\toprule
Contrast & Estimate & SE & df & $t$ & $p$ \\
\midrule
\baseline{} -- \gallery{} & $-23.77$ & $6.69$ & $84$ & $-3.553$ & $\mathbf{0.006}$ \\
\baseline{} -- \handPalette{} & $11.86$ & $6.69$ & $84$ & $1.773$ & $0.799$ \\
\baseline{} -- \headPalette{} & $-5.64$ & $6.69$ & $84$ & $-0.842$ & $1.000$ \\
\baseline{} -- \steeringWheel{} & $-1.55$ & $6.69$ & $84$ & $-0.231$ & $1.000$ \\
\gallery{} -- \handPalette{} & $35.64$ & $6.69$ & $84$ & $5.326$ & $\mathbf{< 0.001}$ \\
\gallery{} -- \headPalette{} & $18.14$ & $6.69$ & $84$ & $2.710$ & $0.081$ \\
\gallery{} -- \steeringWheel{} & $22.23$ & $6.69$ & $84$ & $3.322$ & $\mathbf{0.013}$ \\
\handPalette{} -- \headPalette{} & $-17.50$ & $6.69$ & $84$ & $-2.615$ & $0.106$ \\
\handPalette{} -- \steeringWheel{} & $-13.41$ & $6.69$ & $84$ & $-2.004$ & $0.483$ \\
\headPalette{} -- \steeringWheel{} & $4.09$ & $6.69$ & $84$ & $0.611$ & $1.000$ \\
\bottomrule
\end{tabular}
\end{table}

\begin{table}[t]
\centering
\scriptsize
\setlength{\tabcolsep}{5pt}
\caption{Pairwise post hoc comparisons of \emph{UEQ Hedonic} questionnaire responses across Interfaces (Bonferroni adjustment for 36 tests). $p$-values are bolded if $< 0.05$.}
\begin{tabular}{lrrrrr}
\toprule
Contrast & Estimate & SE & df & $t$ & $p$ \\
\midrule
\baseline{} -- \portalGallery{} & $-69.41$ & $13.40$ & $168$ & $-5.189$ & $\mathbf{< 0.001}$ \\
\baseline{} -- \portalHandPalette{} & $-80.14$ & $13.20$ & $168$ & $-6.050$ & $\mathbf{< 0.001}$ \\
\baseline{} -- \portalHeadPalette{} & $-48.96$ & $13.60$ & $168$ & $-3.613$ & $\mathbf{0.014}$ \\
\baseline{} -- \portalSteeringWheel{} & $-89.36$ & $13.40$ & $168$ & $-6.681$ & $\mathbf{< 0.001}$ \\
\baseline{} -- \wimGallery{} & $-69.73$ & $13.40$ & $168$ & $-5.213$ & $\mathbf{< 0.001}$ \\
\baseline{} -- \wimHandPalette{} & $-75.09$ & $13.40$ & $168$ & $-5.614$ & $\mathbf{< 0.001}$ \\
\baseline{} -- \wimHeadPalette{} & $-60.02$ & $13.40$ & $168$ & $-4.487$ & $\mathbf{0.001}$ \\
\baseline{} -- \wimSteeringWheel{} & $-98.84$ & $13.40$ & $168$ & $-7.389$ & $\mathbf{< 0.001}$ \\
\portalGallery{} -- \portalHandPalette{} & $-10.73$ & $13.20$ & $168$ & $-0.810$ & $1.000$ \\
\portalGallery{} -- \portalHeadPalette{} & $20.45$ & $13.60$ & $168$ & $1.509$ & $1.000$ \\
\portalGallery{} -- \portalSteeringWheel{} & $-19.96$ & $13.40$ & $168$ & $-1.492$ & $1.000$ \\
\portalGallery{} -- \wimGallery{} & $-0.32$ & $13.40$ & $168$ & $-0.024$ & $1.000$ \\
\portalGallery{} -- \wimHandPalette{} & $-5.68$ & $13.40$ & $168$ & $-0.425$ & $1.000$ \\
\portalGallery{} -- \wimHeadPalette{} & $9.39$ & $13.40$ & $168$ & $0.702$ & $1.000$ \\
\portalGallery{} -- \wimSteeringWheel{} & $-29.43$ & $13.40$ & $168$ & $-2.200$ & $1.000$ \\
\portalHandPalette{} -- \portalHeadPalette{} & $31.18$ & $13.50$ & $168$ & $2.317$ & $0.782$ \\
\portalHandPalette{} -- \portalSteeringWheel{} & $-9.22$ & $13.20$ & $168$ & $-0.696$ & $1.000$ \\
\portalHandPalette{} -- \wimGallery{} & $10.41$ & $13.20$ & $168$ & $0.786$ & $1.000$ \\
\portalHandPalette{} -- \wimHandPalette{} & $5.05$ & $13.20$ & $168$ & $0.381$ & $1.000$ \\
\portalHandPalette{} -- \wimHeadPalette{} & $20.12$ & $13.20$ & $168$ & $1.519$ & $1.000$ \\
\portalHandPalette{} -- \wimSteeringWheel{} & $-18.70$ & $13.20$ & $168$ & $-1.412$ & $1.000$ \\
\portalHeadPalette{} -- \portalSteeringWheel{} & $-40.40$ & $13.60$ & $168$ & $-2.981$ & $0.119$ \\
\portalHeadPalette{} -- \wimGallery{} & $-20.76$ & $13.60$ & $168$ & $-1.532$ & $1.000$ \\
\portalHeadPalette{} -- \wimHandPalette{} & $-26.13$ & $13.60$ & $168$ & $-1.928$ & $1.000$ \\
\portalHeadPalette{} -- \wimHeadPalette{} & $-11.06$ & $13.60$ & $168$ & $-0.816$ & $1.000$ \\
\portalHeadPalette{} -- \wimSteeringWheel{} & $-49.88$ & $13.60$ & $168$ & $-3.680$ & $\mathbf{0.011}$ \\
\portalSteeringWheel{} -- \wimGallery{} & $19.64$ & $13.40$ & $168$ & $1.468$ & $1.000$ \\
\portalSteeringWheel{} -- \wimHandPalette{} & $14.27$ & $13.40$ & $168$ & $1.067$ & $1.000$ \\
\portalSteeringWheel{} -- \wimHeadPalette{} & $29.34$ & $13.40$ & $168$ & $2.193$ & $1.000$ \\
\portalSteeringWheel{} -- \wimSteeringWheel{} & $-9.48$ & $13.40$ & $168$ & $-0.708$ & $1.000$ \\
\wimGallery{} -- \wimHandPalette{} & $-5.36$ & $13.40$ & $168$ & $-0.401$ & $1.000$ \\
\wimGallery{} -- \wimHeadPalette{} & $9.71$ & $13.40$ & $168$ & $0.725$ & $1.000$ \\
\wimGallery{} -- \wimSteeringWheel{} & $-29.11$ & $13.40$ & $168$ & $-2.176$ & $1.000$ \\
\wimHandPalette{} -- \wimHeadPalette{} & $15.07$ & $13.40$ & $168$ & $1.126$ & $1.000$ \\
\wimHandPalette{} -- \wimSteeringWheel{} & $-23.75$ & $13.40$ & $168$ & $-1.775$ & $1.000$ \\
\wimHeadPalette{} -- \wimSteeringWheel{} & $-38.82$ & $13.40$ & $168$ & $-2.902$ & $0.151$ \\
\bottomrule
\end{tabular}
\end{table}

\begin{table}[t]
\centering
\scriptsize
\setlength{\tabcolsep}{5pt}
\caption{Pairwise post hoc comparisons of \emph{Continuity} questionnaire responses across Interfaces (Bonferroni adjustment for 36 tests). $p$-values are bolded if $< 0.05$.}
\begin{tabular}{lrrrrr}
\toprule
Contrast & Estimate & SE & df & $t$ & $p$ \\
\midrule
\baseline{} -- \portalGallery{} & $-55.55$ & $14.6$ & $168$ & $-3.816$ & $\mathbf{0.007}$ \\
\baseline{} -- \portalHandPalette{} & $-64.83$ & $14.4$ & $168$ & $-4.498$ & $\mathbf{0.001}$ \\
\baseline{} -- \portalHeadPalette{} & $-40.87$ & $14.7$ & $168$ & $-2.772$ & $0.224$ \\
\baseline{} -- \portalSteeringWheel{} & $-73.34$ & $14.6$ & $168$ & $-5.039$ & $\mathbf{< 0.001}$ \\
\baseline{} -- \wimGallery{} & $-48.30$ & $14.6$ & $168$ & $-3.318$ & $\mathbf{0.040}$ \\
\baseline{} -- \wimHandPalette{} & $-63.77$ & $14.6$ & $168$ & $-4.381$ & $\mathbf{0.001}$ \\
\baseline{} -- \wimHeadPalette{} & $-39.23$ & $14.6$ & $168$ & $-2.695$ & $0.279$ \\
\baseline{} -- \wimSteeringWheel{} & $-74.68$ & $14.6$ & $168$ & $-5.131$ & $\mathbf{< 0.001}$ \\
\portalGallery{} -- \portalHandPalette{} & $-9.29$ & $14.4$ & $168$ & $-0.644$ & $1.000$ \\
\portalGallery{} -- \portalHeadPalette{} & $14.67$ & $14.7$ & $168$ & $0.995$ & $1.000$ \\
\portalGallery{} -- \portalSteeringWheel{} & $-17.80$ & $14.6$ & $168$ & $-1.223$ & $1.000$ \\
\portalGallery{} -- \wimGallery{} & $7.25$ & $14.6$ & $168$ & $0.498$ & $1.000$ \\
\portalGallery{} -- \wimHandPalette{} & $-8.23$ & $14.6$ & $168$ & $-0.565$ & $1.000$ \\
\portalGallery{} -- \wimHeadPalette{} & $16.32$ & $14.6$ & $168$ & $1.121$ & $1.000$ \\
\portalGallery{} -- \wimSteeringWheel{} & $-19.14$ & $14.6$ & $168$ & $-1.315$ & $1.000$ \\
\portalHandPalette{} -- \portalHeadPalette{} & $23.96$ & $14.6$ & $168$ & $1.636$ & $1.000$ \\
\portalHandPalette{} -- \portalSteeringWheel{} & $-8.51$ & $14.4$ & $168$ & $-0.590$ & $1.000$ \\
\portalHandPalette{} -- \wimGallery{} & $16.54$ & $14.4$ & $168$ & $1.147$ & $1.000$ \\
\portalHandPalette{} -- \wimHandPalette{} & $1.06$ & $14.4$ & $168$ & $0.074$ & $1.000$ \\
\portalHandPalette{} -- \wimHeadPalette{} & $25.61$ & $14.4$ & $168$ & $1.777$ & $1.000$ \\
\portalHandPalette{} -- \wimSteeringWheel{} & $-9.85$ & $14.4$ & $168$ & $-0.683$ & $1.000$ \\
\portalHeadPalette{} -- \portalSteeringWheel{} & $-32.47$ & $14.7$ & $168$ & $-2.202$ & $1.000$ \\
\portalHeadPalette{} -- \wimGallery{} & $-7.42$ & $14.7$ & $168$ & $-0.503$ & $1.000$ \\
\portalHeadPalette{} -- \wimHandPalette{} & $-22.90$ & $14.7$ & $168$ & $-1.553$ & $1.000$ \\
\portalHeadPalette{} -- \wimHeadPalette{} & $1.65$ & $14.7$ & $168$ & $0.112$ & $1.000$ \\
\portalHeadPalette{} -- \wimSteeringWheel{} & $-33.81$ & $14.7$ & $168$ & $-2.293$ & $0.832$ \\
\portalSteeringWheel{} -- \wimGallery{} & $25.05$ & $14.6$ & $168$ & $1.721$ & $1.000$ \\
\portalSteeringWheel{} -- \wimHandPalette{} & $9.57$ & $14.6$ & $168$ & $0.657$ & $1.000$ \\
\portalSteeringWheel{} -- \wimHeadPalette{} & $34.11$ & $14.6$ & $168$ & $2.344$ & $0.730$ \\
\portalSteeringWheel{} -- \wimSteeringWheel{} & $-1.34$ & $14.6$ & $168$ & $-0.092$ & $1.000$ \\
\wimGallery{} -- \wimHandPalette{} & $-15.48$ & $14.6$ & $168$ & $-1.063$ & $1.000$ \\
\wimGallery{} -- \wimHeadPalette{} & $9.07$ & $14.6$ & $168$ & $0.623$ & $1.000$ \\
\wimGallery{} -- \wimSteeringWheel{} & $-26.39$ & $14.6$ & $168$ & $-1.813$ & $1.000$ \\
\wimHandPalette{} -- \wimHeadPalette{} & $24.55$ & $14.6$ & $168$ & $1.686$ & $1.000$ \\
\wimHandPalette{} -- \wimSteeringWheel{} & $-10.91$ & $14.6$ & $168$ & $-0.749$ & $1.000$ \\
\wimHeadPalette{} -- \wimSteeringWheel{} & $-35.45$ & $14.6$ & $168$ & $-2.436$ & $0.573$ \\
\bottomrule
\end{tabular}
\end{table}

\begin{table}[t]
\centering
\scriptsize
\setlength{\tabcolsep}{5pt}
\caption{Pairwise post hoc comparisons of \emph{NASA TLX} scores across Interfaces (Bonferroni adjustment for 36 tests). $p$-values are bolded if $< 0.05$.}
\begin{tabular}{lrrrrr}
\toprule
Contrast & Estimate & SE & df & $t$ & $p$ \\
\midrule
\baseline{} -- \portalGallery{} & $-14.75$ & $13.6$ & $168$ & $-1.086$ & $1.000$ \\
\baseline{} -- \portalHandPalette{} & $-3.83$ & $13.5$ & $168$ & $-0.285$ & $1.000$ \\
\baseline{} -- \portalHeadPalette{} & $-44.47$ & $13.8$ & $168$ & $-3.231$ & $0.054$ \\
\baseline{} -- \portalSteeringWheel{} & $-3.98$ & $13.6$ & $168$ & $-0.293$ & $1.000$ \\
\baseline{} -- \wimGallery{} & $2.73$ & $13.6$ & $168$ & $0.201$ & $1.000$ \\
\baseline{} -- \wimHandPalette{} & $24.71$ & $13.6$ & $168$ & $1.818$ & $1.000$ \\
\baseline{} -- \wimHeadPalette{} & $-11.21$ & $13.6$ & $168$ & $-0.825$ & $1.000$ \\
\baseline{} -- \wimSteeringWheel{} & $18.68$ & $13.6$ & $168$ & $1.375$ & $1.000$ \\
\portalGallery{} -- \portalHandPalette{} & $10.92$ & $13.5$ & $168$ & $0.812$ & $1.000$ \\
\portalGallery{} -- \portalHeadPalette{} & $-29.72$ & $13.8$ & $168$ & $-2.159$ & $1.000$ \\
\portalGallery{} -- \portalSteeringWheel{} & $10.77$ & $13.6$ & $168$ & $0.793$ & $1.000$ \\
\portalGallery{} -- \wimGallery{} & $17.48$ & $13.6$ & $168$ & $1.286$ & $1.000$ \\
\portalGallery{} -- \wimHandPalette{} & $39.46$ & $13.6$ & $168$ & $2.904$ & $0.151$ \\
\portalGallery{} -- \wimHeadPalette{} & $3.55$ & $13.6$ & $168$ & $0.261$ & $1.000$ \\
\portalGallery{} -- \wimSteeringWheel{} & $33.43$ & $13.6$ & $168$ & $2.460$ & $0.536$ \\
\portalHandPalette{} -- \portalHeadPalette{} & $-40.65$ & $13.7$ & $168$ & $-2.974$ & $0.122$ \\
\portalHandPalette{} -- \portalSteeringWheel{} & $-0.15$ & $13.5$ & $168$ & $-0.011$ & $1.000$ \\
\portalHandPalette{} -- \wimGallery{} & $6.56$ & $13.5$ & $168$ & $0.487$ & $1.000$ \\
\portalHandPalette{} -- \wimHandPalette{} & $28.53$ & $13.5$ & $168$ & $2.121$ & $1.000$ \\
\portalHandPalette{} -- \wimHeadPalette{} & $-7.38$ & $13.5$ & $168$ & $-0.548$ & $1.000$ \\
\portalHandPalette{} -- \wimSteeringWheel{} & $22.51$ & $13.5$ & $168$ & $1.673$ & $1.000$ \\
\portalHeadPalette{} -- \portalSteeringWheel{} & $40.50$ & $13.8$ & $168$ & $2.942$ & $0.134$ \\
\portalHeadPalette{} -- \wimGallery{} & $47.20$ & $13.8$ & $168$ & $3.429$ & $\mathbf{0.028}$ \\
\portalHeadPalette{} -- \wimHandPalette{} & $69.18$ & $13.8$ & $168$ & $5.025$ & $\mathbf{< 0.001}$ \\
\portalHeadPalette{} -- \wimHeadPalette{} & $33.27$ & $13.8$ & $168$ & $2.417$ & $0.602$ \\
\portalHeadPalette{} -- \wimSteeringWheel{} & $63.16$ & $13.8$ & $168$ & $4.588$ & $\mathbf{< 0.001}$ \\
\portalSteeringWheel{} -- \wimGallery{} & $6.71$ & $13.6$ & $168$ & $0.493$ & $1.000$ \\
\portalSteeringWheel{} -- \wimHandPalette{} & $28.68$ & $13.6$ & $168$ & $2.111$ & $1.000$ \\
\portalSteeringWheel{} -- \wimHeadPalette{} & $-7.23$ & $13.6$ & $168$ & $-0.532$ & $1.000$ \\
\portalSteeringWheel{} -- \wimSteeringWheel{} & $22.66$ & $13.6$ & $168$ & $1.668$ & $1.000$ \\
\wimGallery{} -- \wimHandPalette{} & $21.98$ & $13.6$ & $168$ & $1.617$ & $1.000$ \\
\wimGallery{} -- \wimHeadPalette{} & $-13.93$ & $13.6$ & $168$ & $-1.025$ & $1.000$ \\
\wimGallery{} -- \wimSteeringWheel{} & $15.96$ & $13.6$ & $168$ & $1.174$ & $1.000$ \\
\wimHandPalette{} -- \wimHeadPalette{} & $-35.91$ & $13.6$ & $168$ & $-2.643$ & $0.324$ \\
\wimHandPalette{} -- \wimSteeringWheel{} & $-6.02$ & $13.6$ & $168$ & $-0.443$ & $1.000$ \\
\wimHeadPalette{} -- \wimSteeringWheel{} & $29.89$ & $13.6$ & $168$ & $2.200$ & $1.000$ \\
\bottomrule
\end{tabular}
\end{table}

\begin{table}[h]
\centering
\scriptsize
\setlength{\tabcolsep}{5pt}
\caption{Pairwise post hoc comparisons of \emph{SUS scores} across Interfaces (Bonferroni adjustment for 36 tests). $p$-values are bolded if $< 0.05$.}
\begin{tabular}{lrrrrr}
\toprule
Contrast & Estimate & SE & df & $t$ & $p$ \\
\midrule
\baseline{} -- \portalGallery{} & $34.16$ & $13.9$ & $168$ & $2.460$ & $0.537$ \\
\baseline{} -- \portalHandPalette{} & $22.96$ & $13.8$ & $168$ & $1.670$ & $1.000$ \\
\baseline{} -- \portalHeadPalette{} & $93.54$ & $14.1$ & $168$ & $6.649$ & $\mathbf{< 0.001}$ \\
\baseline{} -- \portalSteeringWheel{} & $19.27$ & $13.9$ & $168$ & $1.388$ & $1.000$ \\
\baseline{} -- \wimGallery{} & $34.96$ & $13.9$ & $168$ & $2.517$ & $0.460$ \\
\baseline{} -- \wimHandPalette{} & $-0.46$ & $13.9$ & $168$ & $-0.033$ & $1.000$ \\
\baseline{} -- \wimHeadPalette{} & $51.30$ & $13.9$ & $168$ & $3.694$ & $\mathbf{0.011}$ \\
\baseline{} -- \wimSteeringWheel{} & $10.73$ & $13.9$ & $168$ & $0.773$ & $1.000$ \\
\portalGallery{} -- \portalHandPalette{} & $-11.20$ & $13.8$ & $168$ & $-0.814$ & $1.000$ \\
\portalGallery{} -- \portalHeadPalette{} & $59.38$ & $14.1$ & $168$ & $4.221$ & $\mathbf{0.001}$ \\
\portalGallery{} -- \portalSteeringWheel{} & $-14.89$ & $13.9$ & $168$ & $-1.072$ & $1.000$ \\
\portalGallery{} -- \wimGallery{} & $0.80$ & $13.9$ & $168$ & $0.057$ & $1.000$ \\
\portalGallery{} -- \wimHandPalette{} & $-34.61$ & $13.9$ & $168$ & $-2.493$ & $0.491$ \\
\portalGallery{} -- \wimHeadPalette{} & $17.14$ & $13.9$ & $168$ & $1.234$ & $1.000$ \\
\portalGallery{} -- \wimSteeringWheel{} & $-23.43$ & $13.9$ & $168$ & $-1.687$ & $1.000$ \\
\portalHandPalette{} -- \portalHeadPalette{} & $70.58$ & $14.0$ & $168$ & $5.053$ & $\mathbf{< 0.001}$ \\
\portalHandPalette{} -- \portalSteeringWheel{} & $-3.69$ & $13.8$ & $168$ & $-0.268$ & $1.000$ \\
\portalHandPalette{} -- \wimGallery{} & $11.99$ & $13.8$ & $168$ & $0.872$ & $1.000$ \\
\portalHandPalette{} -- \wimHandPalette{} & $-23.42$ & $13.8$ & $168$ & $-1.703$ & $1.000$ \\
\portalHandPalette{} -- \wimHeadPalette{} & $28.34$ & $13.8$ & $168$ & $2.061$ & $1.000$ \\
\portalHandPalette{} -- \wimSteeringWheel{} & $-12.23$ & $13.8$ & $168$ & $-0.890$ & $1.000$ \\
\portalHeadPalette{} -- \portalSteeringWheel{} & $-74.27$ & $14.1$ & $168$ & $-5.279$ & $\mathbf{< 0.001}$ \\
\portalHeadPalette{} -- \wimGallery{} & $-58.59$ & $14.1$ & $168$ & $-4.164$ & $\mathbf{0.002}$ \\
\portalHeadPalette{} -- \wimHandPalette{} & $-93.99$ & $14.1$ & $168$ & $-6.681$ & $\mathbf{< 0.001}$ \\
\portalHeadPalette{} -- \wimHeadPalette{} & $-42.25$ & $14.1$ & $168$ & $-3.003$ & $0.111$ \\
\portalHeadPalette{} -- \wimSteeringWheel{} & $-82.82$ & $14.1$ & $168$ & $-5.886$ & $\mathbf{< 0.001}$ \\
\portalSteeringWheel{} -- \wimGallery{} & $15.68$ & $13.9$ & $168$ & $1.129$ & $1.000$ \\
\portalSteeringWheel{} -- \wimHandPalette{} & $-19.73$ & $13.9$ & $168$ & $-1.421$ & $1.000$ \\
\portalSteeringWheel{} -- \wimHeadPalette{} & $32.02$ & $13.9$ & $168$ & $2.306$ & $0.804$ \\
\portalSteeringWheel{} -- \wimSteeringWheel{} & $-8.55$ & $13.9$ & $168$ & $-0.615$ & $1.000$ \\
\wimGallery{} -- \wimHandPalette{} & $-35.41$ & $13.9$ & $168$ & $-2.550$ & $0.420$ \\
\wimGallery{} -- \wimHeadPalette{} & $16.34$ & $13.9$ & $168$ & $1.177$ & $1.000$ \\
\wimGallery{} -- \wimSteeringWheel{} & $-24.23$ & $13.9$ & $168$ & $-1.745$ & $1.000$ \\
\wimHandPalette{} -- \wimHeadPalette{} & $51.75$ & $13.9$ & $168$ & $3.727$ & $\mathbf{0.010}$ \\
\wimHandPalette{} -- \wimSteeringWheel{} & $11.18$ & $13.9$ & $168$ & $0.805$ & $1.000$ \\
\wimHeadPalette{} -- \wimSteeringWheel{} & $-40.57$ & $13.9$ & $168$ & $-2.921$ & $0.143$ \\
\bottomrule
\end{tabular}
\end{table}

\clearpage

\subsection{Performance Across All Trials}
\begin{figure}[h]
    \centering
    \includegraphics[width=\linewidth]{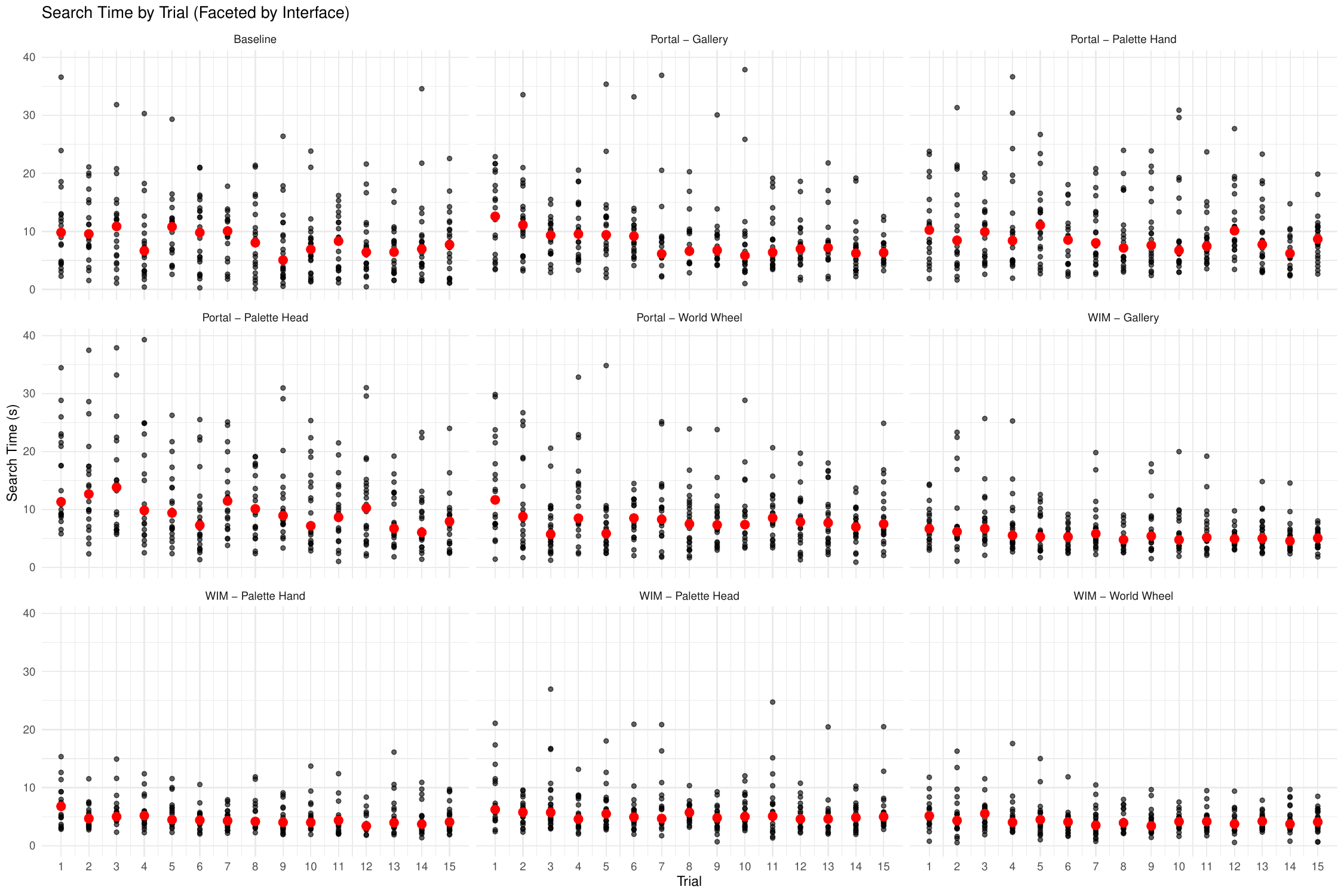}
    \caption{Search Time by Trial per Interface. Individual task times are shown as semi-transparent points, while red markers indicate the 
median task time for each trial. The plots are faceted by Interface.}
    \label{fig:searchbytrial}
\end{figure}

\begin{figure}[h]
    \centering
    \includegraphics[width=\linewidth]{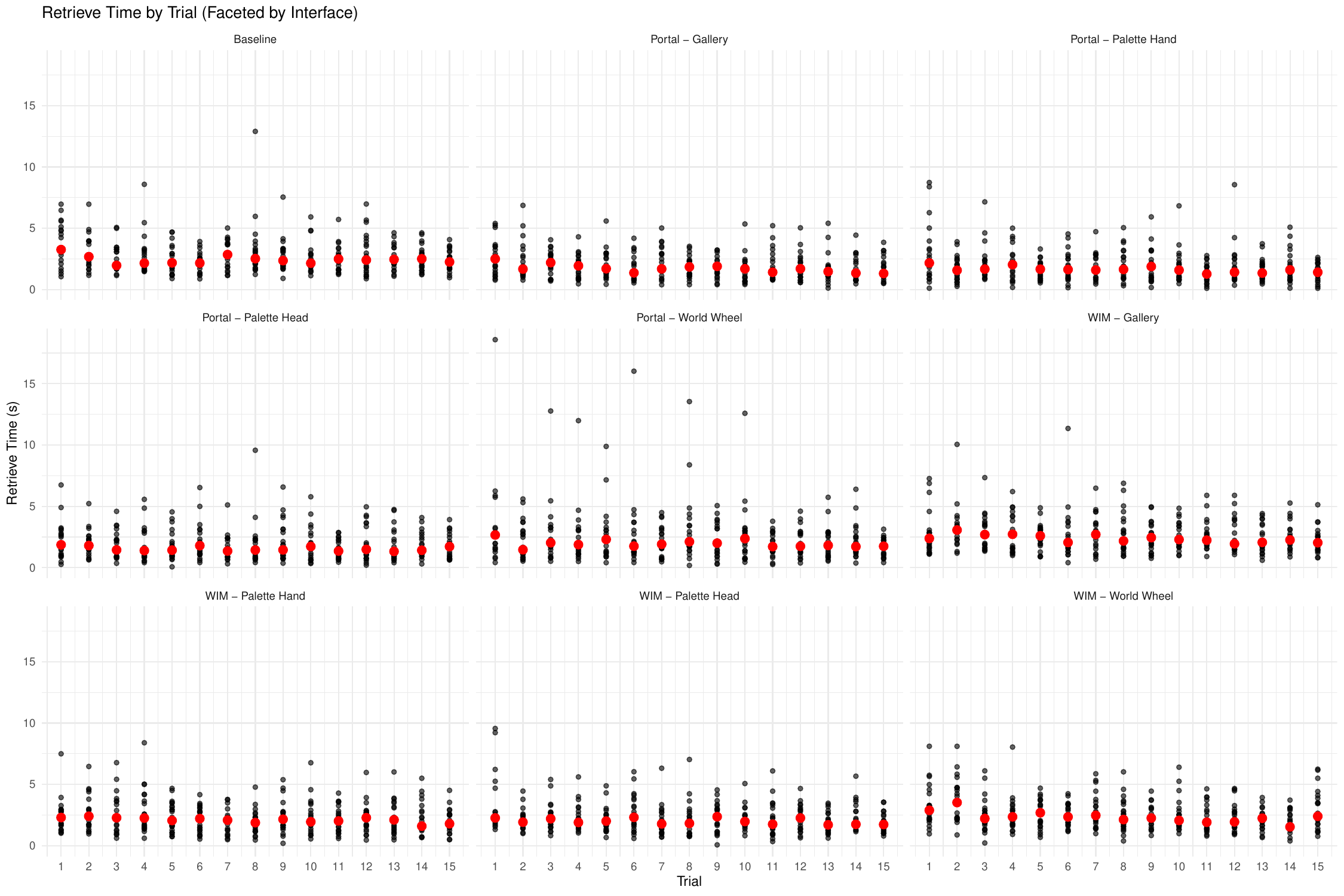}
    \caption{Retrieve Time by Trial per Interface. Individual task times are shown as semi-transparent points, while red markers indicate the 
median task time for each trial. The plots are faceted by Interface.}
    \label{fig:retrievebytrail}
\end{figure}

\begin{figure}[h]
    \centering
    \includegraphics[width=\linewidth]{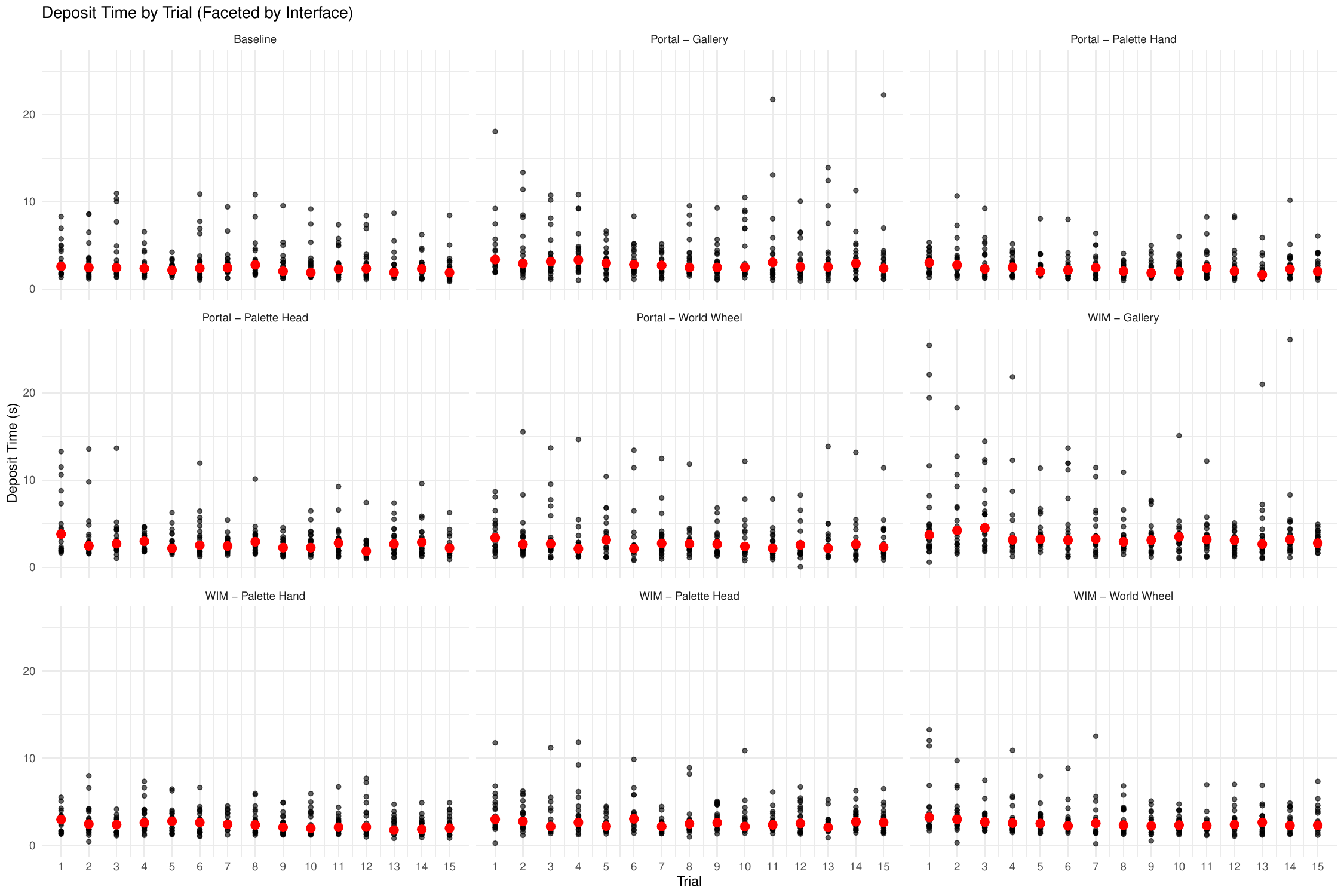}
    \caption{Deposit Time by Trial per Interface. Individual task times are shown as semi-transparent points, while red markers indicate the 
median task time for each trial. The plots are faceted by Interface.}
    \label{fig:depositbytrial}
\end{figure}

\end{document}